\documentclass[useAMS,usenatbib]{mnras}

\usepackage[english]{babel}
\usepackage[fleqn]{amsmath}
\usepackage{amsfonts}
\usepackage{txfonts}
\usepackage{graphicx}
\usepackage{amssymb}
\usepackage[T1]{fontenc}
\usepackage{ae,aecompl}
\usepackage{fancyhdr}
\usepackage{multicol}
\usepackage{layout}
\usepackage{graphicx}
\usepackage{multirow}
\usepackage{color}
\usepackage{multirow}
\usepackage{times}
\usepackage[outdir=./]{epstopdf}

\newif\ifAMStwofonts
\AMStwofontstrue

\newcommand{\ltsima}{$\; \buildrel < \over \sim \;$}
\newcommand{\lsim}{\lower.5ex\hbox{\ltsima}}
\newcommand{\gtsima}{$\; \buildrel > \over \sim \;$}
\newcommand{\gsim}{\lower.5ex\hbox{\gtsima}}
\newcommand{\bra}{\left\langle}
\newcommand{\ket}{\right\rangle}

\newcommand{\dd}{\mathrm{d}}

\newcommand{\ci}{\mathrm{i}}

\newcommand{\dirac}{\delta_D}

\title[Variations of covariance matrices]{Variations of cosmic large-scale structure covariance matrices across parameter space}
\author[R. Reischke, A. Kiessling and B.M. Sch{\"a}fer]{Robert Reischke$^{1}$\thanks{E-mail:
reischke@stud.uni-heidelberg.de}, Alina Kiessling$^{2}$ and Bj\"orn Malte Sch\"afer$^{1}$\\
$^{1}$Zentrum f\"ur Astronomie der Universit{\"a}t Heidelberg, Astronomisches Recheninstitut, Philosophenweg 12, D-69120, Heidelberg, Germany \\
$^{2}$Jet Propulsion Laboratory, California Institute of Technology, 4800 Oak Grove Drive, Pasadena, California, United States of America}

\begin{document}

\date{}

\pagerange{\pageref{firstpage}--\pageref{lastpage}} \pubyear{}
\maketitle

\label{firstpage}

\begin{abstract}
The likelihood function for cosmological parameters, given by e.g. weak lensing shear measurements, depends on contributions to the covariance induced by the nonlinear evolution of the cosmic web. As nonlinear clustering to date has only been described by numerical $N$-body simulations in a reliable and sufficiently precise way, the necessary computational costs for estimating those covariances at different points in parameter space are tremendous. In this work we describe the change of the matter covariance and of the weak lensing covariance matrix as a function of cosmological parameters by constructing a suitable basis, where we model the contribution to the covariance from nonlinear structure formation using Eulerian perturbation theory at third order. We show that our formalism is capable of dealing with large matrices and reproduces expected degeneracies and scaling with cosmological parameters in a reliable way. Comparing our analytical results to numerical simulations we find that the method describes the variation of the covariance matrix found in the SUNGLASS weak lensing simulation pipeline within the errors at one-loop and tree-level for the spectrum and the trispectrum, respectively, for multipoles up to $\ell\leq 1300$. We show that it is possible to optimize the sampling of parameter space where numerical simulations should be carried out by minimising interpolation errors and propose a corresponding method to distribute points in parameter space in an economical way.
\end{abstract}

\begin{keywords}
gravitational lensing: weak, dark energy, large-scale structure of Universe.
\end{keywords}

\section{Introduction}\label{sec:1}
Measurements of cosmological parameters and investigations into the properties of gravity on large scales are the focus of a number of upcoming surveys of the cosmic large-scale structure. These investigations require probing how the expansion dynamics of the Universe and the gravitational model affect the growth rate of structures, as well as understanding the relation between redshift and distance. A tool combining both these sources of cosmological information is weak gravitational lensing \citep[e.g.][]{Kaiser1998, Bacon2000, Kaiser2000, Maoli2000, Mellier2000, Bartelmann2001, Kilbinger2003} which, as a line-of-sight integrated quantity of the Newtonian tidal shear field, probes both structure growth and the evolution of the background cosmology by measuring a correlation in the shapes of galaxies.

The estimation of cosmological parameters based on large-scale structure observations requires a precise knowledge of the covariance matrix, which describes the cosmic variance, the statistical dependence of the modes of the cosmic matter distribution, and the noise inherent in the surveys. Due to mode coupling in non-linear structure formation the covariance matrix is non-diagonal, acquires large amplitudes on small scales, and renders the statistical properties of the cosmic matter distribution non-Gaussian; In this respect cosmological large-scale structure observations differ significantly from observations of primary covariance matrix-fluctuations \citep{Komatsu2011}, where the assumption of Gaussian statistics is very good. 

The scaling of non-linear structure growth with cosmological parameters is necessarily non-linear, which is immediately apparent in perturbative approaches \citep{Bernardeau1994, Bernardeau1994a, Taruya2002}. Each order of perturbation theory is characterised by a different dependence on cosmology, and in assembling a perturbation series these dependences are mixed by superposition. Mode coupling in non-linear structure formation generates off-diagonal entries in the covariance matrix \citep[e.g.][etc.]{Scoccimarro1999a, Cooray2001, Takada2007, Takada2009, Sato2009, Kayo2012} and therefore reduces the information content of the density field \citep[e.g.][]{Hu2003, Takada2007,Sato2009, Sato2011}. On the other hand, fluctuations in the cosmic density field are strongly amplified by non-linear structure formation, which allows measurements on small scales which are otherwise inaccessible due to the sparsity of galaxies. Future experiments such as the Euclid mission\footnote{http://www.euclid-ec.org} \citep{Laureijs2011} will use the weak gravitational lensing effect to probe the cosmic web on scales deep in the non-linear regime \citep[e.g.][]{Benjamin2007, Laureijs2011, VanWaerbeke2013, Kitching2014}. In fact, Euclid's anticipated weak lensing signal, with a significance of close to $1000\sigma$, is largely generated by non-linear scales.

Because non-linear structure formation cannot yet	 be fully described by analytical methods, estimates of the covariance matrix require simulations of cosmic structure formation. Due to the large volume of future surveys and the necessity to observe at non-linear scales, cosmological simulations require both large volumes and high resolutions. In addition, a large suite of statistically equivalent simulations is required to estimate covariance matrices using ensemble-averaging. This estimation needs to be undertaken throughout the anticipated parameter space, because non-linear structure formation depends strongly on the choice of cosmological parameters.

Standard spatially flat dark energy cosmologies typically have six parameters. Thus, even a rather coarse sampling of parameter space would require a tremendous number of $N$-body ray-tracing simulations \citep{Fosalba2008, Hilbert2009} or other techniques such as line-of-sight integrations \citep{Kiessling2011}, which are, up to now, the only robust method to determine the mode coupling and induced higher order cumulants to the desired accuracy. The computational load to produce large suites of simulations at Gpc-scales, while retaining resolution at sub-Mpc-scales, quickly becomes prohibitive. As a consequence, it is inevitable, that variations in the covariance matrices in parameter space are being investigated \citep{Eifler2009} and a way to interpolate between these points must now be developed.

In this context a number of questions arise: $(i)$ How strong are the variations of the covariance matrix with varying cosmological parameters? $(ii)$ To which cosmological parameters is the covariance matrix most sensitive? $(iii)$ Is there a way of predicting variations and in which directions in parameter space the strongest variations are encountered? $(iv)$ Is it possible to decompose changes to the shape, size and orientation of the covariance matrix in a geometrically clear way? $(v)$ What would be sensible choices of cosmological parameters for simulations in order to cover the relevant parameter space economically? $(vi)$ Is there a natural way to interpolate between covariance matrices from numerical simulations?

Recently \citet{Schafer2016} introduced a method to interpolate between Fisher matrices at different points in parameter space We now intend to apply this formalism to the variation of the covariance matrix of the matter and convergence power spectrum estimators. This should be possible because both Fisher-matrices and covariance matrices share positive-definiteness as a common property, which is required by our formalism. The focus will be on the power spectrum of the weak lensing convergence, as it is directly linked to observables provided by Euclid. Non-linear structure formation on small scales generates a non-Gaussian contribution to the covariance matrix, where we employ Eulerian perturbation theory at tree-level to predict the trispectrum as the lowest-order non-Gaussian contribution \citep{Scoccimarro1999a}. We consider perturbation theory as an easily manageable tool for predicting non-linear corrections to the covariance matrix and do not imply that it describes all non-linearities accurately, but we will check its validity against numerical simulations.

The fiducial cosmological model is a spatially flat $\Lambda$CDM model with base parameters
$\Omega_\text{m} = 0.3$, $\Omega_\Lambda = 1-\Omega_\text{m}$, $h = 0.7$, $n_\text{s} =1$, $w_0=-1$ and $w_\text{a} = 0$. Moreover,  we will use the sum convention throughout this paper, thus implying summation over repeated indices. After a brief review of the lensing observables we will review the covariance matrix theory for the matter and convergence spectrum in Sect. \ref{sec:2}. In Sect. \ref{sec:3} the Lie basis is constructed and applied to the covariance matrix in Sect. \ref{sec:4}, where we also compare the theoretical prediction with simulations. In Sect. \ref{sec:5} we summarize.

\section{Covariance Matrices}\label{sec:2}

\subsection{Covariance matrix of the matter spectrum estimator}
A Fourier mode of the density contrast $\delta(\bmath{x})$ is given by
\begin{equation}\label{eq:1}
\delta(\bmath{k}) = \int_{\mathbb R ^3}\dd^3x\:\delta(\bmath{x})\exp(-\text{i}\bmath{k}\bmath{x}).
\end{equation}
Whereas the matter spectrum, $P(k)$, follows ideally in the ensemble-average for a Gaussian homogeneous field, $\bra\delta(\bmath k)\delta(\bmath k^\prime)^*\ket = (2\pi)^3\dirac(\bmath k-\bmath k^\prime)P(k)$, the estimation of $P(k)$ from a survey of finite volume involves cosmic variance and correlations between estimates due to non-linear structure formation.

$P(k)$ can be estimated from a survey of volume $V$ by dividing it into $N$ spherical shells in Fourier-space with radii $k_i$ and width $\Delta k_i$ as the variance of all modes within a shell \citep{Scoccimarro1999a},
\begin{equation}\label{eq:2}
\hat P(k_i) = \frac{1}{V}\int_{k_i}\frac{\dd^3k}{V_\text{s}(k_i)}\:\delta(\bmath k)\delta(-\bmath k),
\end{equation}
where $V_\mathrm{s}(k_i)$ is the volume of the $i$th shell. The Fourier-transform is Hermitean, $\delta(\bmath k) = \delta(-\bmath k)^*$, because $\delta$ is real-valued.

The covariance matrix for the estimates $\hat P(k_i)$ is now given by
\begin{equation}\label{eq:3}
C^\delta_{ij} =  \frac{1}{V}\left[\frac{(2\pi)^3}{V_{\text{s}}(k_i)}2P^2(k_i)\delta_{ij} +\bar T_{ij}\right],
\end{equation}
with a Gaussian, diagonal part, and a non-Gaussian contribution $\bar T_{ij}$,
\begin{equation}\label{eq:4}
\bar T_{ij} = \int_{k_i}\!\!\frac{\dd^3\! k_1}{V_\text{s}(k_i)} \int_{k_j}\!\!\frac{\dd^3\! k_2}{V_\text{s}(k_j)}T(\bmath{k_1},-\bmath{k_1},\bmath{k_2},-\bmath{k_2}),
\end{equation}
related to the matter trispectrum $T$ which appears as the connected part of a 4-point correlation function that does not separate into squares of the matter spectrum if $\delta$ assumes non-Gaussian statistical properties. The emergence of non-Gaussian terms like the matter trispectrum can be approximated by perturbation theory and ultimately require direct simulation.

\subsection{Covariance matrix of the weak lensing spectrum}
The statistical properties of the weak lensing signal are inherited from those of the density field because weak lensing is a linear mapping of density. Consequently, the covariance of the weak lensing power spectrum estimates from a survey will involve a non-Gaussian contribution due to the non-Gaussianity of the underlying density field. Working with the weak lensing convergence, $\kappa_m$, derived from the lensing signal of all galaxies within a tomographic redshift bin $m$,
\begin{equation}\label{eq:5}
\kappa_m(\bmath{\ell}) = 
\int_{\mathbb R ^2}\dd\theta\:\kappa_m(\bmath{\theta})\exp(-\ci\bmath{\ell}\bmath{\theta}),
\end{equation}
where $\kappa_m(\bmath{\theta})$ is given by
\begin{equation}\label{eq:6}
\kappa_m(\bmath{\theta}) = \int_0^{\chi_\text{H}}\dd\chi\:W_m(\chi)\frac{\delta(\chi,\chi\bmath\theta)}{a},
\end{equation}
with the lensing efficiency function $W_m(\chi)$
\begin{equation}\label{eq:8}
W_m(\chi) = \frac{3\Omega_m}{2\chi_H^2}\int_\chi^\infty \frac{\dd z}{\dd\chi^\prime}G_m(\chi^\prime)\left(1-\frac{\chi}{\chi^\prime}\right),
\end{equation}
and $G_m(\chi)$ being the distance distribution of sources in the $m$th bin which is normalised to one, $\int \dd\chi\:G_m(\chi) = 1$.

\begin{figure}
\begin{center}
\includegraphics[width = 0.45\textwidth]{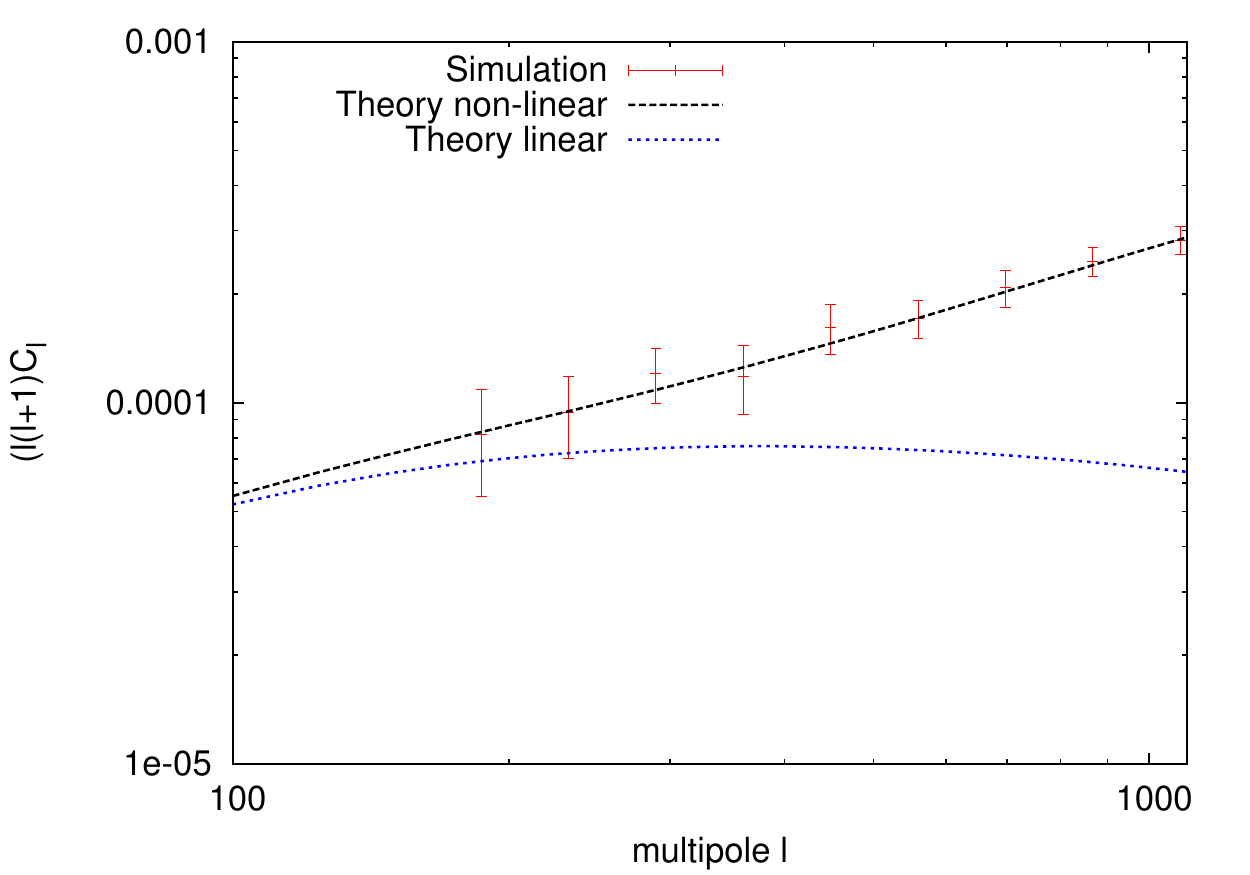}
\caption{Comparison of the convergence power spectrum obtained from the simulation with the linear and non-linear theoretical prediction. The latter is obtained from the fitting formula described in \citet{Smith2003}. Note that the errors shown for the simulation are taken to be the diagonal parts of the covariance matrix. Therefore they are not uncorrelated.}
\label{Fig:SimPow}
\end{center}
\end{figure}

The convergence power spectrum is now given as
\begin{equation}\label{eq:9}
C^\kappa_{mn}(\ell)= \int\frac{\dd\chi}{\chi^2} W_m(\chi)W_n(\chi)\:P(\ell/\chi,\chi),
\end{equation}
while the projected trispectrum of the weak lensing convergence is given by
\begin{equation}\label{eq:10}
T^\kappa_{mn} =  
\int\frac{\dd\chi}{\chi^6}\:W_m^2(\chi)W_n^2(\chi)\:T(\bmath\ell_1/\chi,-\bmath\ell_1/\chi,\bmath\ell_2/\chi,-\bmath\ell_2/\chi),
\end{equation}
where $T^\kappa_{mn}$ depends on the combination $(\bmath\ell_1,-\bmath\ell_1,\bmath\ell_2,-\bmath\ell_2)$ of wave vectors.

The covariance of estimates $\hat C^\kappa_{ij}(\ell)$ of the tomographic weak lensing power spectra proceeds in complete analogy to the previous case: The solid angle $\Omega$ in Fourier-space is divided into $N$ rings centered at the wave vectors $\ell_m$, with width $\Delta \ell_m$ and volume $A_r(\ell_m)$. The covariance matrix of the estimates $\hat C^\kappa_{ij}(\ell_m)$ is then given by
\begin{equation}\label{eq:11}
C^\kappa_{ij,mn} = 
\frac{1}{\Omega}\left(2\:C^\kappa_{mn}(\ell_m)^2\frac{(2\pi)^2}{A_r(\ell_m)}\delta_{ij} + T^\kappa_{ij,mn}\right),
\end{equation}
with
\begin{equation}\label{eq:12}
T^\kappa_{ij,mn} = \int_{\ell_1}\frac{\dd\ell_1}{A_r(\ell_n)}\int_{\ell_2}\frac{\dd\ell_2}{A_r(\ell_n)}
T^\kappa_{mn}(\bmath\ell_1,-\bmath\ell_1,\bmath\ell_2,-\bmath\ell_2).
\end{equation}
Note that the diagonal elements contain, in principle, the shot noise term due to the finite number of background galaxies and their intrinsic ellipticity distribution. This term, however, does not depend on cosmology and is neglected in our analysis. Naturally, the covariance matrices will depend on the cosmological model and will undergo a transformation if a cosmological parameter assumes a new value; Effectively, we will require a set of transformation matrices for each direction of the parameter space, which is provided exactly by our formalism.

\begin{figure*}
\begin{center}
\includegraphics[width = 0.45\textwidth]{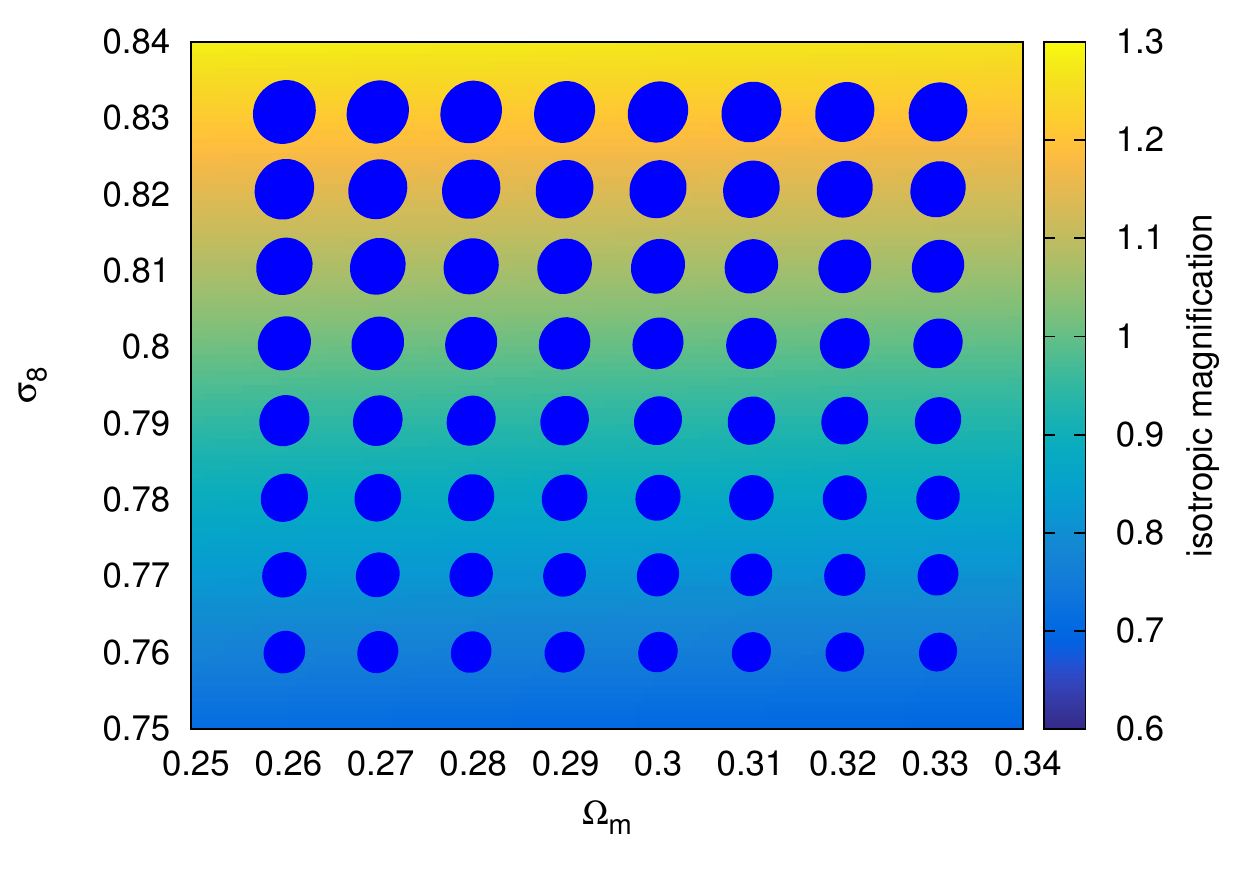}
\includegraphics[width = 0.45\textwidth]{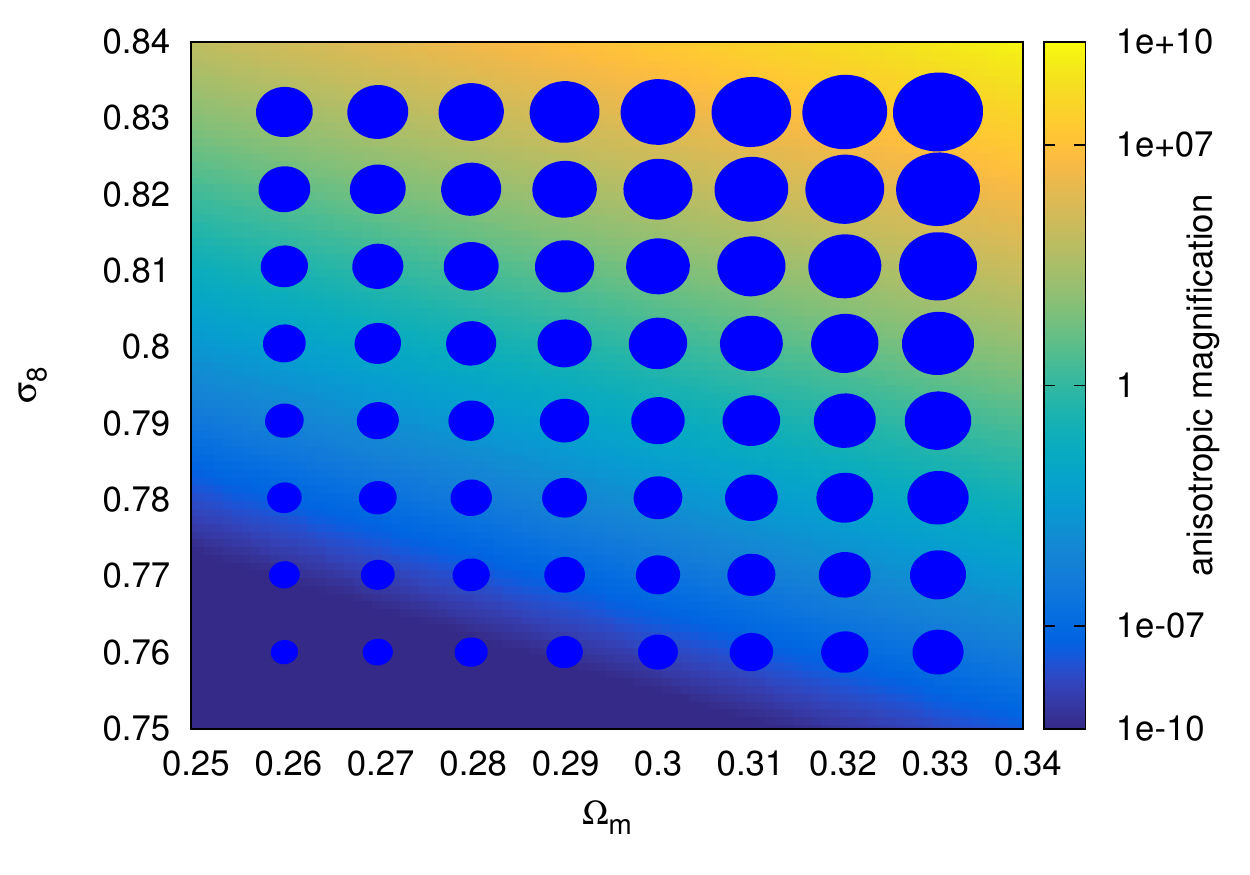}
\caption{Variations of the matter spectrum covariance matrix in the $\Omega_\text{m}-\sigma_8$ plane. \textit{Left}: We show the trace of the covariance matrix as isotropic magnification. Furthermore, we show the quadratic form induced by two $k$-bins as ellipses. The $k$-bin combination is $(k_1,k_2)$. \textit{Right}: The determinant of the covariance matrix is shown as anisotropic magnification. $k$-bins are chosen to be $(k_N,k_{N-1})$.}
\label{Fig:1}
\end{center}
\end{figure*}

\begin{figure*}
\begin{center}
\includegraphics[width = 0.45\textwidth]{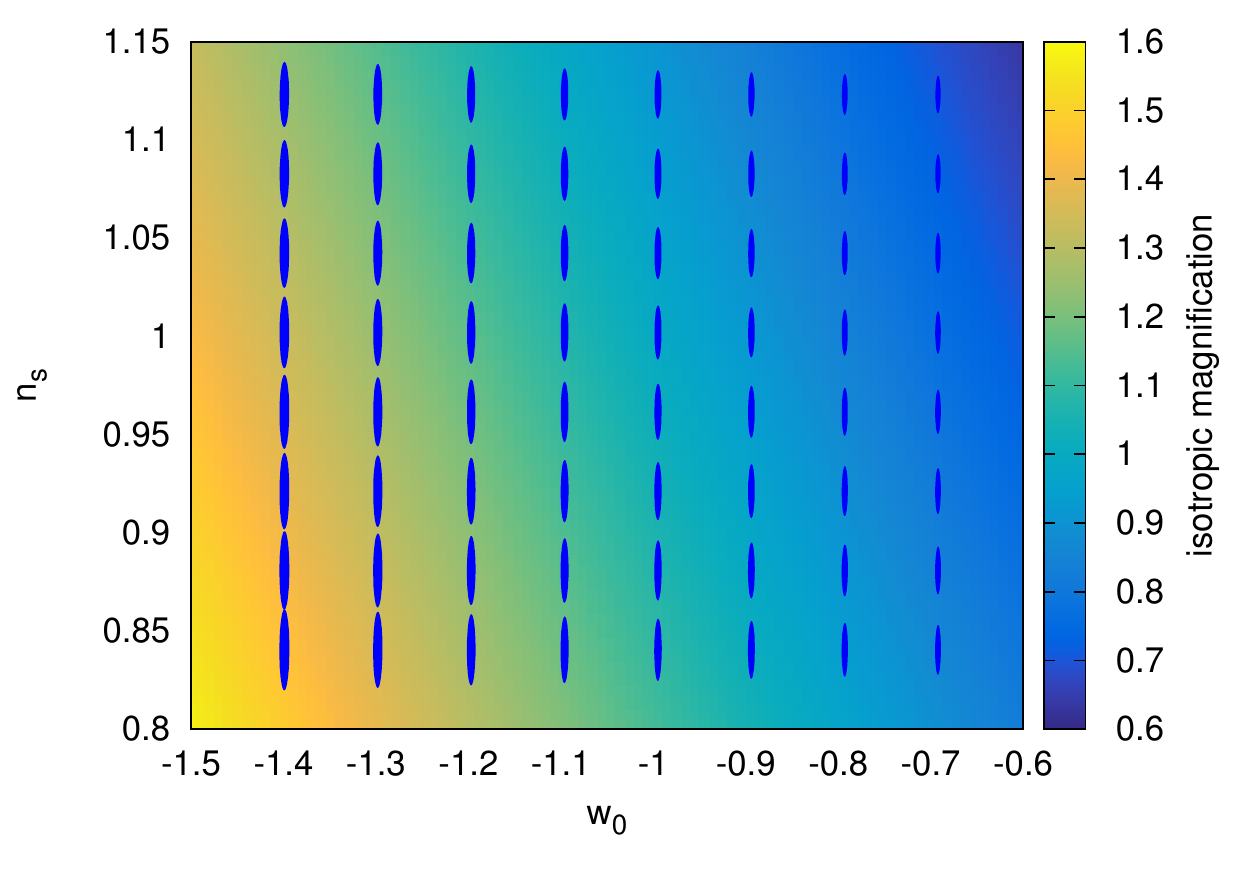}
\includegraphics[width = 0.45\textwidth]{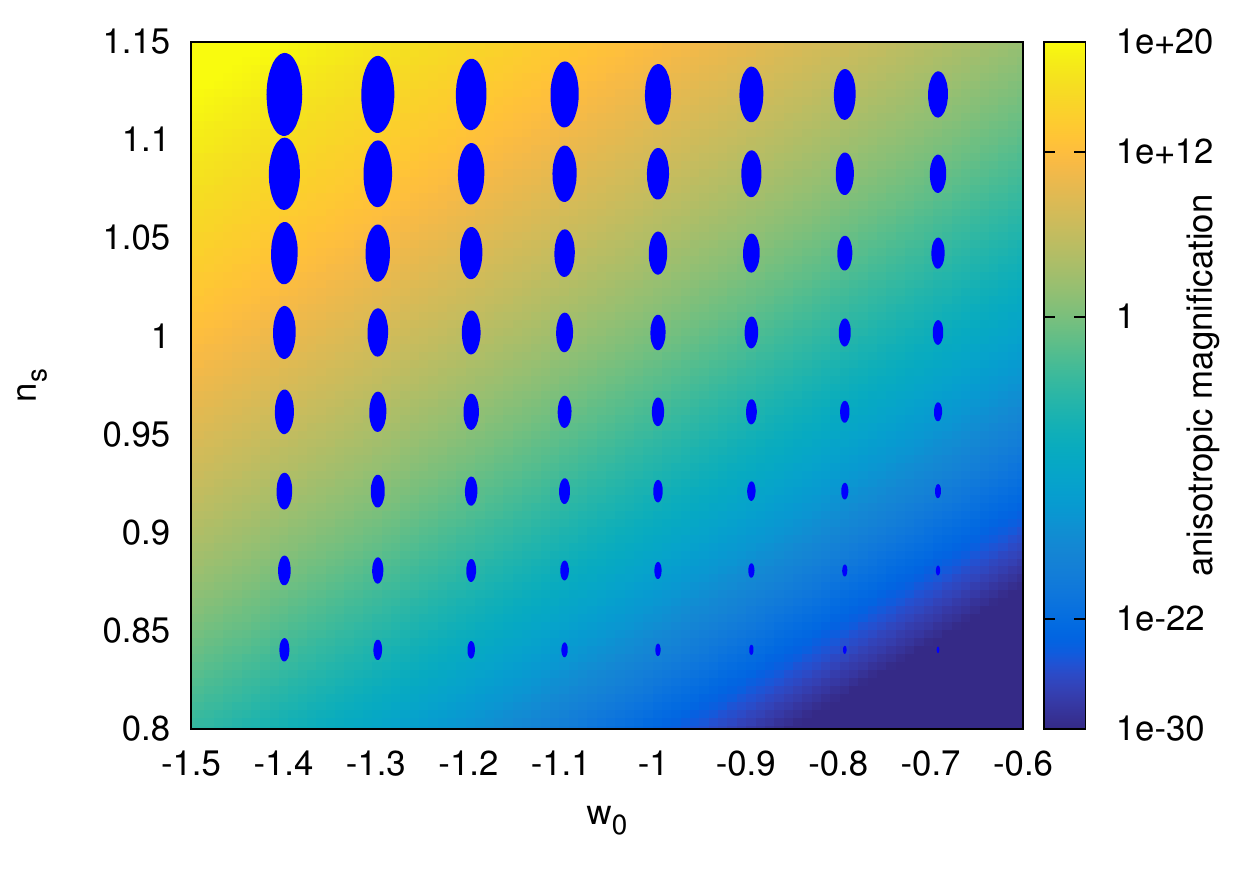}
\includegraphics[width = 0.45\textwidth]{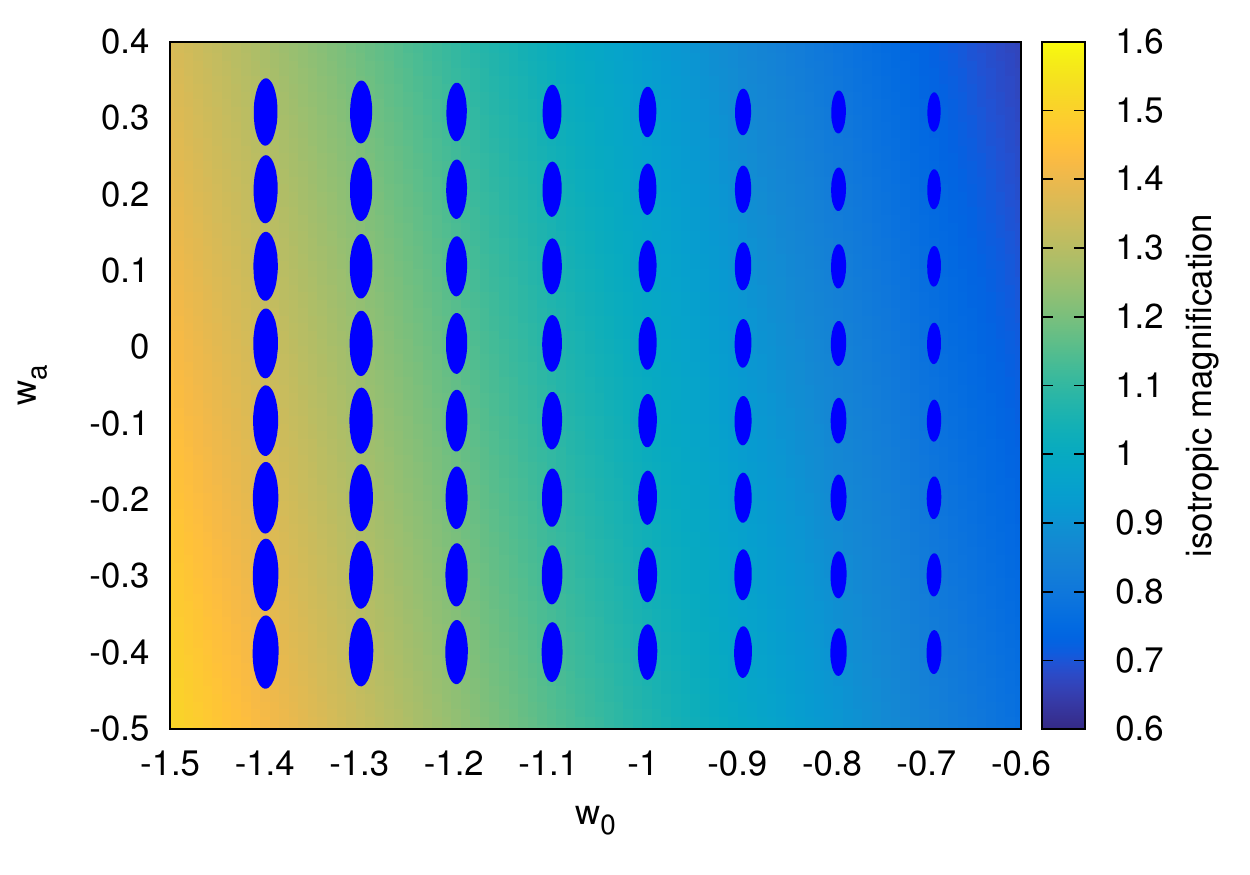}
\includegraphics[width = 0.45\textwidth]{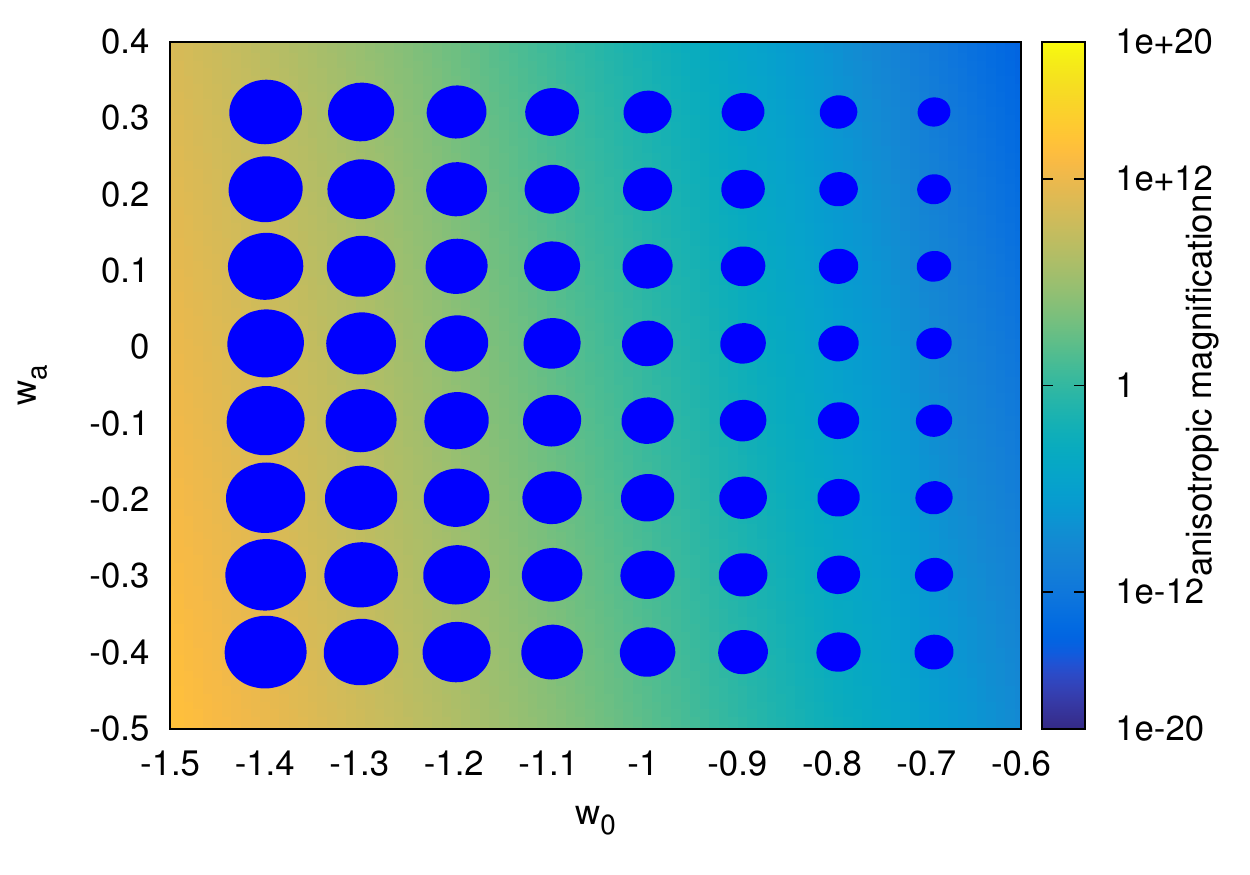}
\includegraphics[width = 0.45\textwidth]{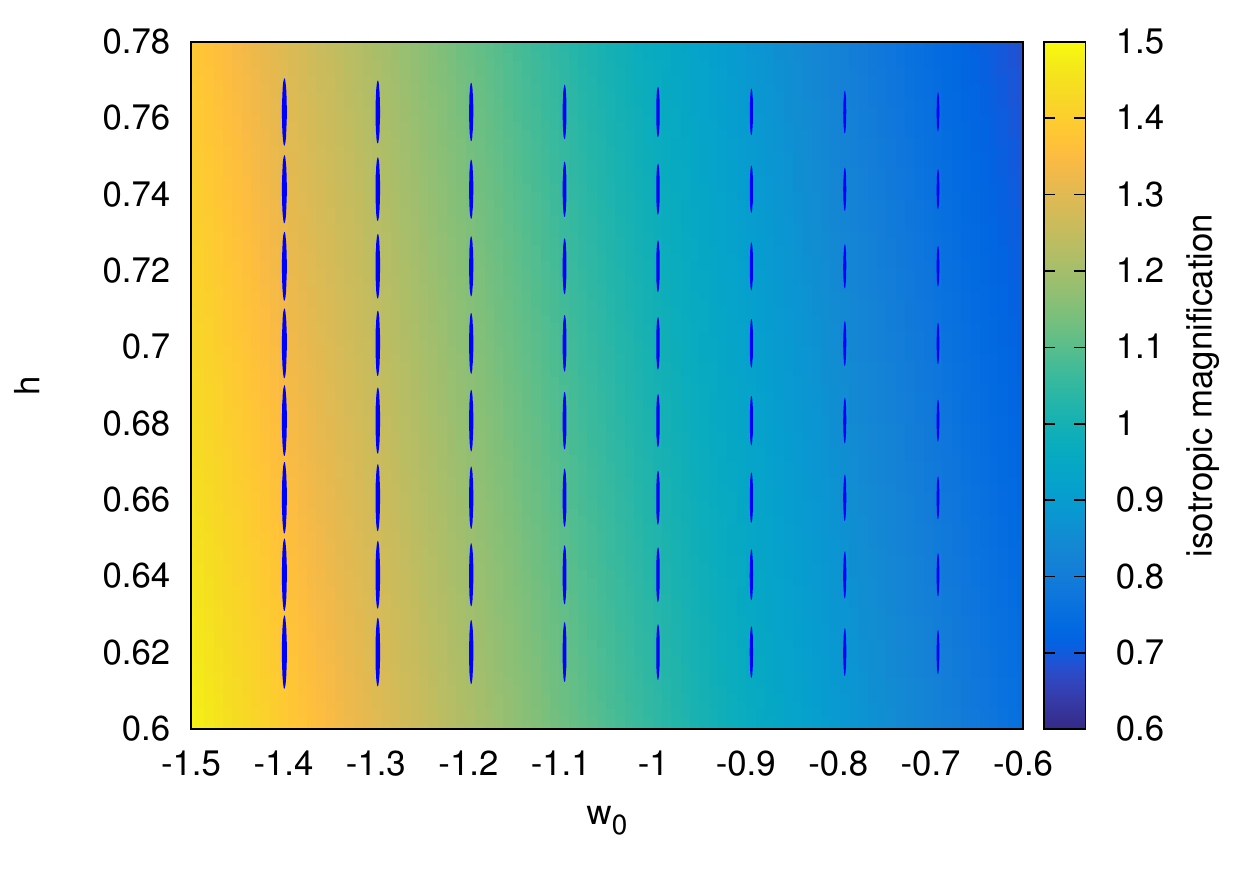}
\includegraphics[width = 0.45\textwidth]{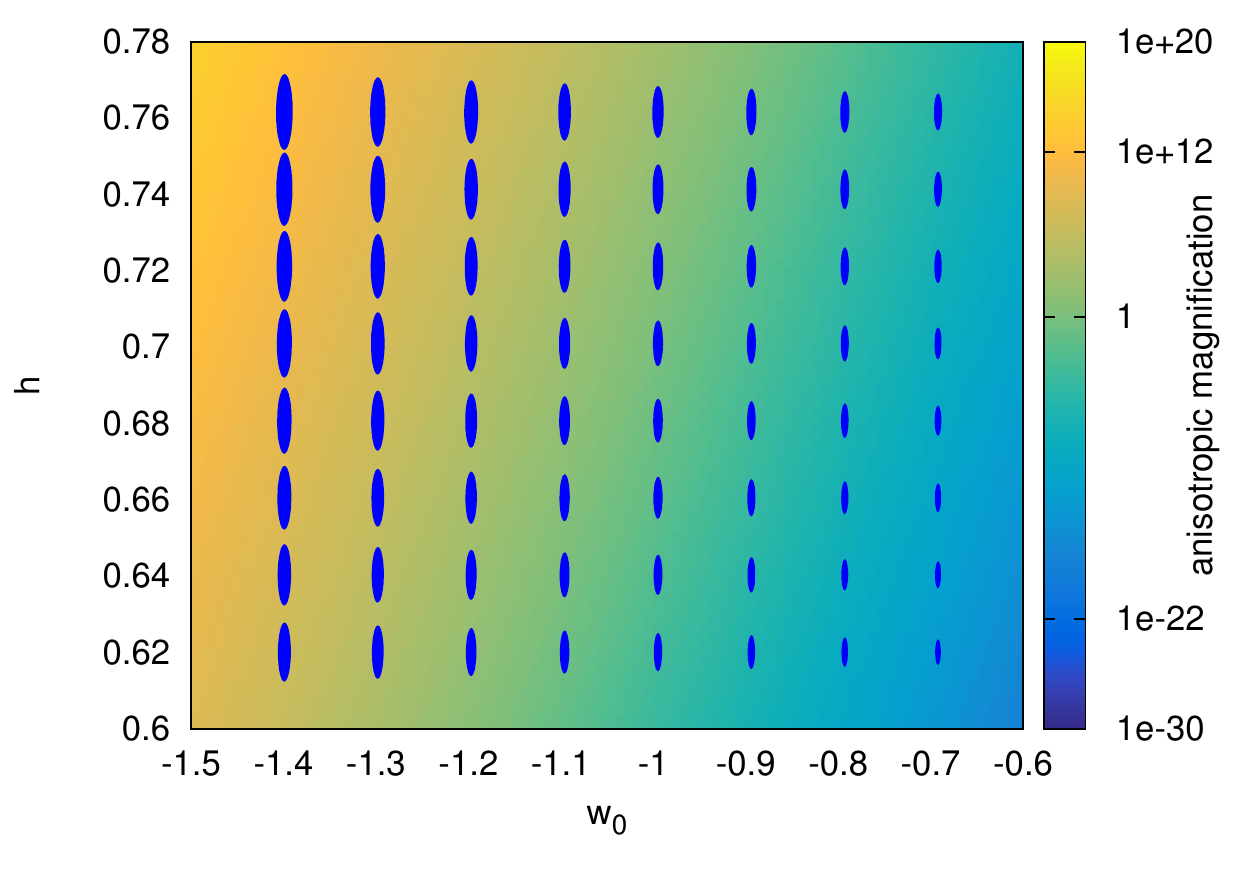}
\includegraphics[width = 0.45\textwidth]{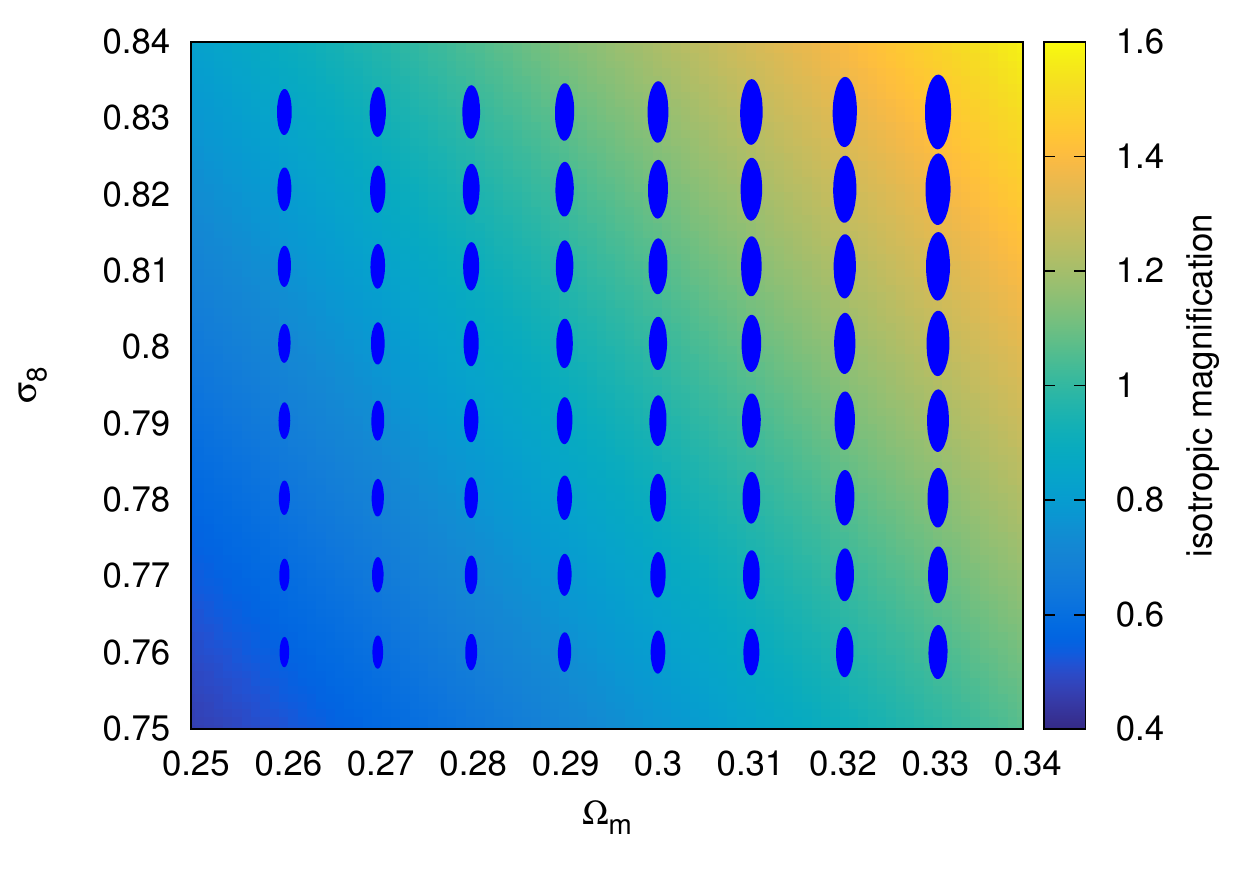}
\includegraphics[width = 0.45\textwidth]{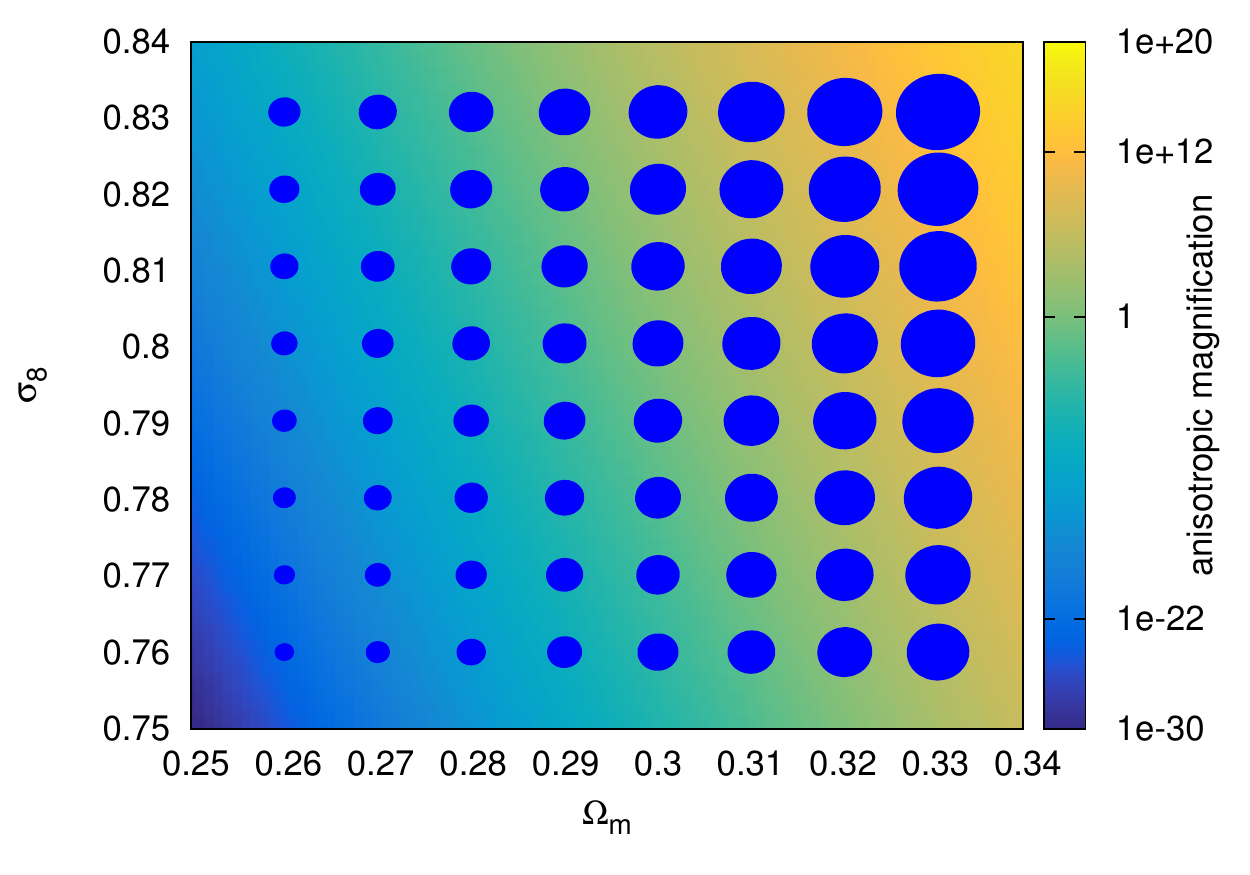}
\caption{Variations of the convergence spectrum covariance matrix in the different parameter planes. \textit{Left}: We show the trace of the covariance matrix as isotropic magnification. Furthermore we show the quadratic form induced by two $\ell$-bins as ellipses. The $\ell$-bin combination is $(\ell_1,\ell_2)$. \textit{Right}: The determinant of the covariance matrix is shown as anisotropic magnification. $\ell$-bins are chosen to be $(\ell_N,\ell_{N-1})$.}
\label{Fig:2}
\end{center}
\end{figure*}

\subsection{Trispectrum at tree-level}
The covariance matrix is diagonal in the limit of linear structure formation which conserves the Gaussianity of the initial conditions and the independence of the Fourier-modes. In this case, the 4-point correlator arising in the expression for the covariance separates into squares of spectra by virtue of Wick's theorem. This is different in non-linear gravitational clustering, where mode coupling renders the statistical properties of the density field non-Gaussian and generates a trispectrum contribution to the covariance matrix. There exist various different approaches to non-linear structure formation. Eulerian perturbation \citep[e.g.][]{Bernardeau2002} theory operates on a fluid based picture, while Lagrangian perturbation theory \citep[e.g.][]{ZelDovich1970,Buchert1992,Bouchet1995} perturbs initial particle positions and their trajectories. Recently an approach using classical statistical field theory was introduced \citep{Bartelmann2016}. Furthermore, there are more phenomenological models such as the halo model \citep{Cooray2002} or on the empirical log-normal distribution of the density field \citep{2011A&A...536A..85H}.

In this work we use Eulerian perturbation theory as a model for non-Gaussianities in structure formation, since the covariance matrix can be calculated rather easily and it does not depend on additional parameters such as, for example, the halo model. Thus the dependence on cosmological parameters, which we investigate here, enters directly into the perturbative expansion of the linear solution.  We will use third-order perturbations at tree-level because it is an easily manageable model. In Sect. \ref{sec:sim} we will show tests of the accuracy of the model against non-Gaussian lensing convergence maps derived from numerical simulations.

Eulerian perturbation theory, with perturbations of second and third order to the density and velocity fields, gives the trispectrum  expression \citep{Fry1984}
\begin{equation}
\begin{split}
T(\bmath{k_1}, \bmath{k_2},& \bmath{k_3}, \bmath{k_4}) = 4\big[F_2(\bmath{k_{12}},-\bmath{k_1})F_2(\bmath{k_{12}},\bmath{k_3})P_1P_{12}P_3 \\ & + \text{cycl.}\big] + 6\big[F_3(\bmath{k_1},\bmath{k_2}, \bmath{k_3})P_1P_2P_3 +\text{cycl.}\big],
\end{split}
\end{equation}
where we abbreviated $\bmath{k_{12}} \equiv \bmath{k_1} +\bmath{k_2}$, and $P_\text{lin}(k_i) \equiv P_i$ is the linear power spectrum. The latter is given by the usual expression
\begin{equation}
P_\text{lin}(k,a) = D_+^2(a)T^2(k)P_\text{ini}(k),
\end{equation}
where $D_+(a)$ is the normalized growth factor, $T(k)$ is the transfer function from \citet{Bardeen1986}, and $P_\text{ini}(k)$ is the initial power spectrum which is set by inflation to be proportional to $k^{n_\text{s}}$ with the spectral index, $n_\text{s}$, being very close to unity.
 Evaluating the general expression for the configuration of the wave vectors needed in Eq. (\ref{eq:4}) yields
\begin{equation}\begin{split}
T(\bmath{k_1}&,-\bmath{k_1},\bmath{k_2},-\bmath{k_2}) =  12F_3(\bmath{k_1},-\bmath{k_1},\bmath{k_2})P_1^2P_2\\ +& \, 8F_2^2(\bmath{k_1}-\bmath{k_2},\bmath{k_2})P(|\bmath{k_1}-\bmath{k_2}|)P_2^2\\ +& \, 16F_2(\bmath{k_1}-\bmath{k_2},\bmath{k_2})F_2(\bmath{k_2}-\bmath{k_1},\bmath{k_1})P_1P_2P(|\bmath{k_1}-\bmath{k_2}|)\\  +& \,  (\bmath{k_1}\leftrightarrow \bmath{k_2}),
\end{split}
\end{equation}
where the symbol $(\bmath{k_1}\leftrightarrow \bmath{k_2})$ implies a repetition of the previous term with the wave vectors interchanged.

For practical calculations we note that the trispectrum is only a function of the magnitudes and the relative orientation of $\bmath k_1$ and $\bmath k_2$. In order to be consistent in perturbation theory we evaluate the Gaussian term in Eq. (\ref{eq:11}) at the one-loop level given by
\begin{equation}
P(k,t) = P_\text{lin}(k,t) + P_{22}(k,t) + P_{13}(k,t),
\end{equation}
where the two one-loop contributions can be written in terms of the linear spectrum $P_\text{lin}$,
\begin{equation}\begin{split}
P_{22}(k,t) = & \ 2 \int \dd^3q\:\left[F_2(\bmath{k}-\bmath{q},q)\right]^2 P_\text{lin}(|\bmath{k}-\bmath{q}|,t)P_\text{lin}(q,t), \\
P_{13}(k,t) = & \ 6\int\dd^3q\:F_3(\bmath{k},\bmath{q},-\bmath{q})P_\text{lin}(k,t)P_\text{lin}(q,t).
\end{split}\end{equation}
Here it should be noted that although the contributions to the spectrum grow homogeneous in time for each order separately, their linear combination does not because the growth rates of each term are different.

\subsection{Simulations}\label{sec:24}
To estimate the variations of the covariance matrix in comparison to our perturbative model we use simulations generated using the SUNGLASS weak lensing simulation pipeline \citep{Kiessling2011}, which we will briefly summarize here. The pipeline uses the \textsc{Gadget2} \citep{Springel2005} $N$-body code to generate non-linearly evolved cosmic density fields. Specifically, the simulations assume a $\Lambda$CDM cosmology and comprise $512^3$ particles in a simulation box with a side length of $512h^{-1}\text{Mpc}$. Weak lensing shear and convergence maps are derived from simulation snap-shots by carrying out light-of-sight integrations of tidal shear fields under the Born-approximation. These simulated weak lensing light cones cover a solid angle of $100$ square degrees with a depth of $0\leq z\leq 2$ in redshift \citep{Kiessling2011}. The light-cones have a \textit{Euclid}-like source redshift distribution \citep{Refregier2004} with the functional form
\begin{equation}
n(z) \propto z^\alpha\exp\left[-\left(\frac{z}{z_0}\right)^\beta\right]\; .
\end{equation}
The functional form of the source redshift distribution is fitted to the simulation yielding $z_0\approx 0.9$.
In \autoref{Fig:SimPow} we show the convergence power spectrum obtained from the simulation together with the non-linear \citet{Smith2003} and the linear power spectrum. The binning which was chosen for the power spectrum will also be used in Sect. \ref{sec:sim}. Clearly we are dealing with scales which reach deep into the non-linear regime.

Statistically equivalent simulations for a range of choices of $\Omega_\text{m0}$ and $\sigma_8$ are available, summarised by Table~\ref{tab:sim}, which allows the determination of weak lensing covariances as an ensemble average over the weak lensing spectra derived from each simulated map. Averaging over all realizations of each parameter set allows to calculate the covariance of the spectrum estimator via
\begin{equation}\label{eq:CovSim}
C^{\kappa}_{mn} = \left\langle \left(C^\kappa(\ell_m) -\hat C^{\kappa}(\ell_m)\right) \left(C^{\kappa}(\ell_n) -\hat C^\kappa(\ell_n)\right)\right\rangle \; ,
\end{equation} 
where $\hat C^\kappa(\ell_m)$ is the estimated spectrum averaged over all realizations. 

\begin{table}\label{tab:sim}
 \caption{Cosmological parameters of the simulations and the number of realizations}
 \begin{center}
  \begin{tabular}{c|c|c|c|c|c|c}
   \hline
   \hline
   $\Omega_\text{m0}$ & $\Omega_\Lambda$ & $\Omega_\text{b}$ & $h$ &$\sigma_8$ & $n_\text{s}$ & $N_\text{real}$ \\
   0.272 & 0.728 & 0.0449 & 0.71 & 0.809 & 1 & 50 \\
   0.272 & 0.728 & 0.0449 & 0.71 & 0.728 & 1 & 50 \\
   0.272 & 0.728 & 0.0449 & 0.71 & 0.890 & 1 & 50 \\
   0.299 & 0.701 & 0.0449 & 0.71 & 0.809 & 1 & 50 \\
   0.245 & 0.755 & 0.0449 & 0.71 & 0.809 & 1 & 50 \\
   \hline
  \end{tabular}
 \end{center}
\end{table}

\section{Lie Basis}\label{sec:3}
The covariance matrix depends strongly on the choice of the cosmological model; As a quantity involving second powers of spectra in the linear regime and third powers of the spectra in the perturbative non-linear regime, it scales $\propto\sigma_8^{4\ldots6}$. The proportionality of the weak lensing signal with $\Omega_\text{m}$ generates a  dependence $\propto\Omega_\text{m}^4$ and the exact shape of the spectra encapsulated in $n_s$ and $h$ matters due to the mode coupling which determines the superposition of spectra in the expression for the trispectrum. In addition, the weak lensing effect depends on the dark energy properties through the relation between redshift and comoving distance as well as on the amplitude of cosmic structures as a function of distance or redshift. Furthermore, there are degeneracies between the parameters and situations where different parameter choices result in very similar covariance matrices. In summary, small changes in these physical properties account for a variation of the covariance matrix, which illustrates the necessity of accurate models.

Estimates of the covariance matrix require suites of cosmological simulations to be run throughout the expected parameter space, but coverage with a fine grid quickly becomes unfeasible given the dimensionality of basic $w$CDM-models. However, given an understanding of the variations of the covariance matrix, one could distribute the simulations in an economic way by identifying directions of rapid changes of the covariance matrix while sampling the parameter space only sparsely in directions with parameter degeneracies. The starting point of such a description of the variations of the covariance matrix is the construction of a basis, which determines the rate of change with each cosmological parameter.

We follow the procedure outlined in \citet{Schafer2016} describing the change of the covariance matrix $C_{ij}^{\kappa}$ at some fiducial model $x_\alpha$ to another point in parameter space $x^\prime_\alpha$ by the action of a linear transformation $U_{ij}$
\begin{equation}\label{eq:linmap}
C^\prime_{ij} \equiv C_{ij}(x^\prime_\alpha) = U_{ik}C_{km}(x_\alpha)U_{mj},\quad i,j=1,...,N,
\end{equation}
with the dimensionality $N$ of the covariance matrix, i.e. the number of $k$- or $\ell$-bins, which are indexed by $i$ and $j$ in the above formula. For simplicity, we revert to a non-tomographic weak lensing measurement; However, in principle, the tomographic weak lensing spectra would only add a technical complication to the formalism.

We construct the transformation by drawing the matrix root $C_{ij}(x^\prime_\alpha) = B_{ik}B_{kj}$ and identifying identical pairs. Then,
\begin{equation}
U_{ij} = B_{il}(B^{-1})_{lj}.
\end{equation} 
The infinitesimal transformation takes the usual form
\begin{equation}
U_{ij} = \delta_{ij} + (x^\prime_\alpha-x_\alpha)T_{ij\alpha},\quad l=1\ldots M,
\end{equation}
with $T_{ij\alpha}$ being the generators of the transformation. Note that $T_{ij\alpha}$ is a collection of $M$ matrices, i.e. one $N\times N$ matrix for every parameter direction. This is very similar to the action of the connection coefficients in general relativity. $T_{ij\alpha}$ is given by differentiation
\begin{equation}
T_{ij\alpha} = \partial_\alpha U_{ij}.
\end{equation}
Multiple actions of the infinitesimal transformation lead in the limit to the global transformation, which is the usual matrix exponential
\begin{equation}
U_{ij} = \exp\left((x^\prime_\alpha-x_\alpha)T_{ij\alpha}\right).
\end{equation}
Approximating this up to linear order in $x^\prime_\alpha-x_\alpha$ yields
\begin{equation}\label{eq:20}
U_{ij} = \delta_{ij} + (x^\prime_\alpha-x_\alpha)T_{ij\alpha}.
\end{equation}
Note that with this approximation the transformations in different parameter directions commute. Numerically, the generators are derived using finite differencing of $U_{ij}$
\begin{equation}\label{eq:21}
T_{ij\alpha} =\frac{U_{ij}(x_\alpha+\Delta x_\alpha) - U_{ij}(x_\alpha -\Delta x_\alpha)}{2\Delta x_\alpha}.
\end{equation}
In this way it is possible to describe the transformation of the covariance matrix $C_{ij}$ between $\bmath x_\alpha$ and $\bmath x^\prime_\alpha$, and to decompose the transformation $U_{ik}$ in terms of geometrically easy to interpret modes.

\section{Variations of the Covariance Matrices}\label{sec:4}

\subsection{Matter spectrum covariance}
As a proof of concept we calculate the covariance matrix in $N=60$ equidistant $k$-bins of width $\Delta k \approx 0.01\,h\text{Mpc}^{-1}$ with $k_1\equiv k_\text{min} = 0.19\,h\text{Mpc}^{-1}$ and $k_N \equiv k_\text{max} = 0.8\,h\text{Mpc}^{-1}$. We take the survey volume to be unity as it only yields an overall factor.
The covariance matrix is calculated at the fiducial model and at two other points in parameter space for each parameter direction. The variation $\Delta x_\alpha$ for the finite differencing in Eq. (\ref{eq:21}) is chosen to be $0.01$ for $\sigma_8$ and $\Omega_\text{m}$. We show the change of the covariance matrix using the trace, the determinant, and by picking out pairs of $k$-bins, which give rise to a quadratic form which can be represented as an ellipse.

The trace quantifies the isotropic magnification (relative to the fiducial model) of the covariance matrix because the off-diagonal elements do not enter. Conversely, the determinant quantifies the anisotropic magnification due to the fact that the value of the determinant depends on the magnitude of the off-diagonal elements in relation to the diagonal elements. Because the magnitude of the off-diagonal elements is bounded by the geometric mean of the corresponding diagonal elements as a consequence of the Cauchy-Schwarz inequality, large correlation coefficients describe strong degeneracies and therefore strong anisotropic magnification.

\autoref{Fig:1} shows the trace and determinant of the submatrix of the matter spectrum covariance as measures of isotropic and anisotropic magnification. It can be seen that changes in the traces are mainly due to $\sigma_8$ which is expected as it is mainly a rescaling of the elements of the covariance matrix. The dependence of the trace on $\Omega_\text{m}$ is weaker because the matter content influences the shape of the matter spectrum but not the amplitude. In contrast, the determinant shows a degeneracy between $\Omega_\text{m}$ and $\sigma_8$, which is due to the fact that an increasing $\Omega_\text{m}$ shifts the peak of the spectrum to higher $k$ values, thus the mode coupling terms in the trispectrum include different values at different values of $k$.

Blue ellipses indicate the magnitude and the correlation coefficient of the covariance matrix in two different $k$ bins. In the left plot we show the covariance matrix for the wave vector pair $(k_1,k_2)$ while the right plot shows the covariance matrix for the pair $(k_{N-1},k_N)$. Clearly the behaviour of the ellipses follows the behaviour of the magnifications. This also shows that at low $k$ the shape of the covariance matrix is dominated by $\sigma_8$ and quantifies linearly evolving scales due to the near diagonality, while at higher $k$ also the off-diagonal elements become important as a consequence of non-linear structure formation.

\begin{figure*}
\begin{center}
\includegraphics[width = 0.45\textwidth]{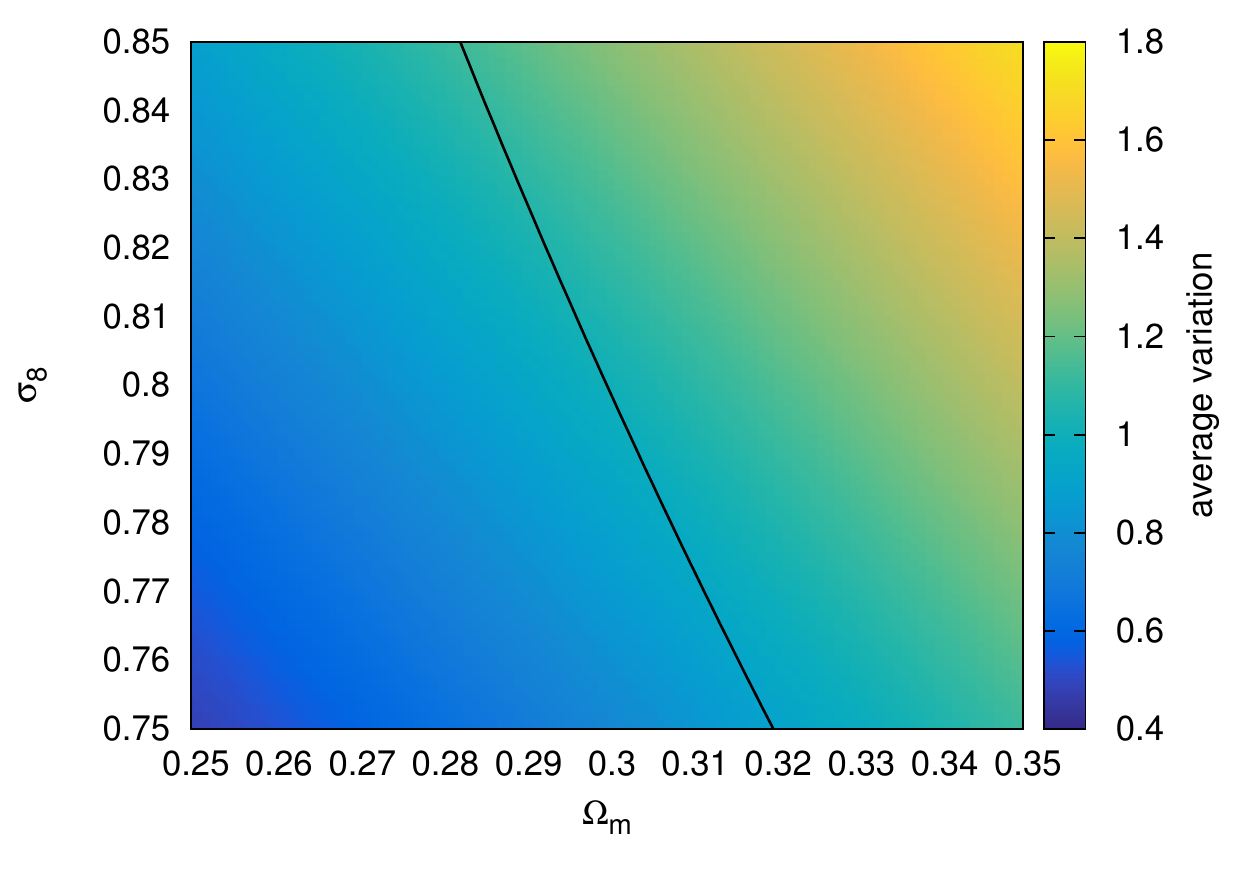}
\includegraphics[width = 0.45\textwidth]{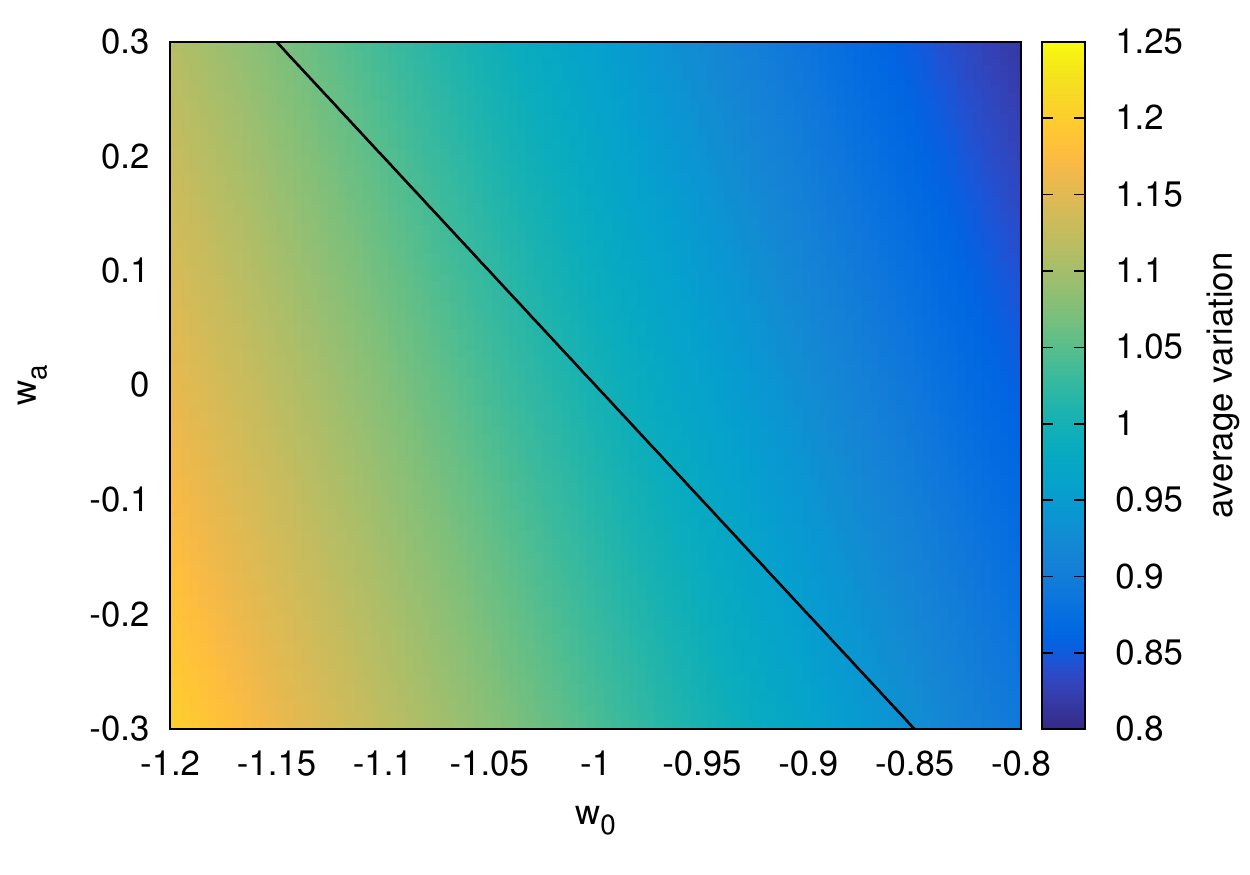}
\includegraphics[width = 0.45\textwidth]{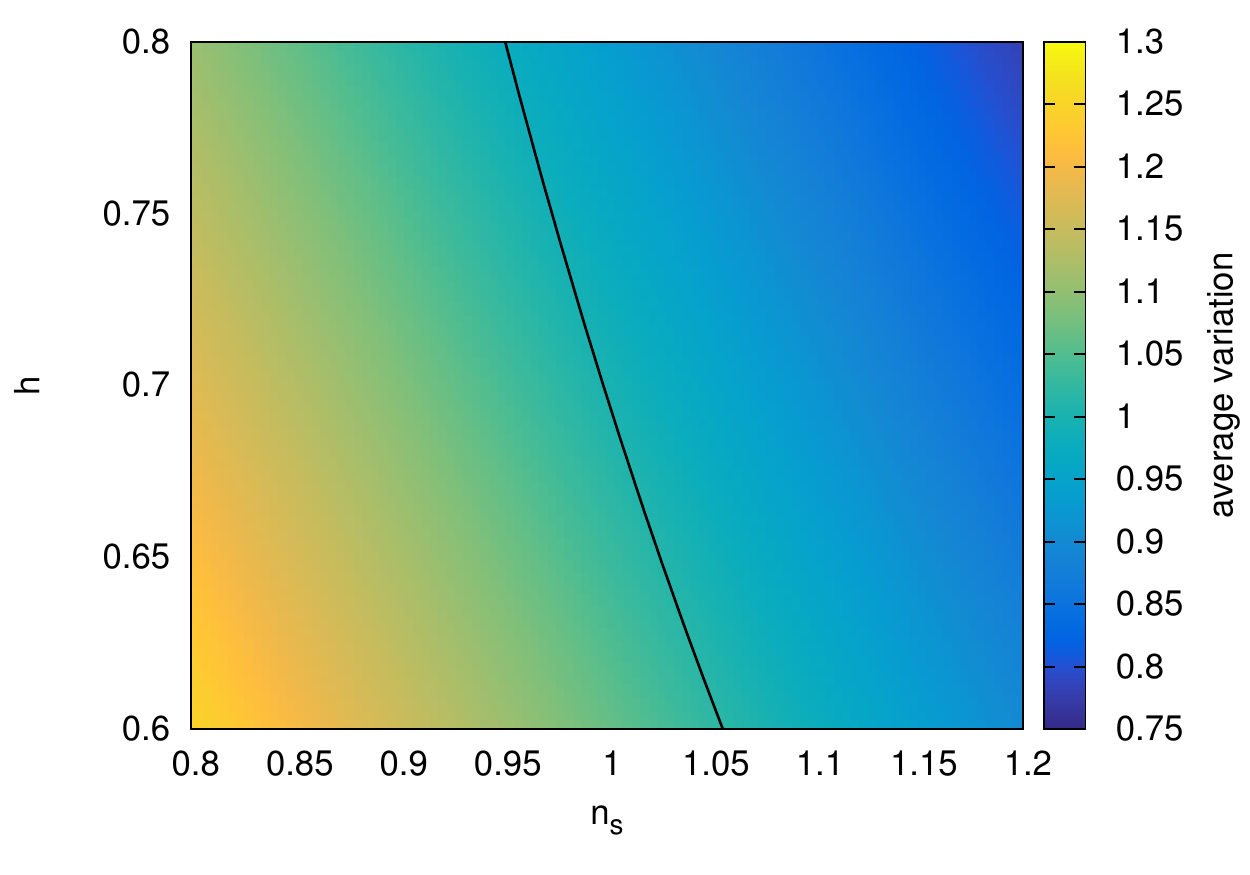}
\includegraphics[width = 0.45\textwidth]{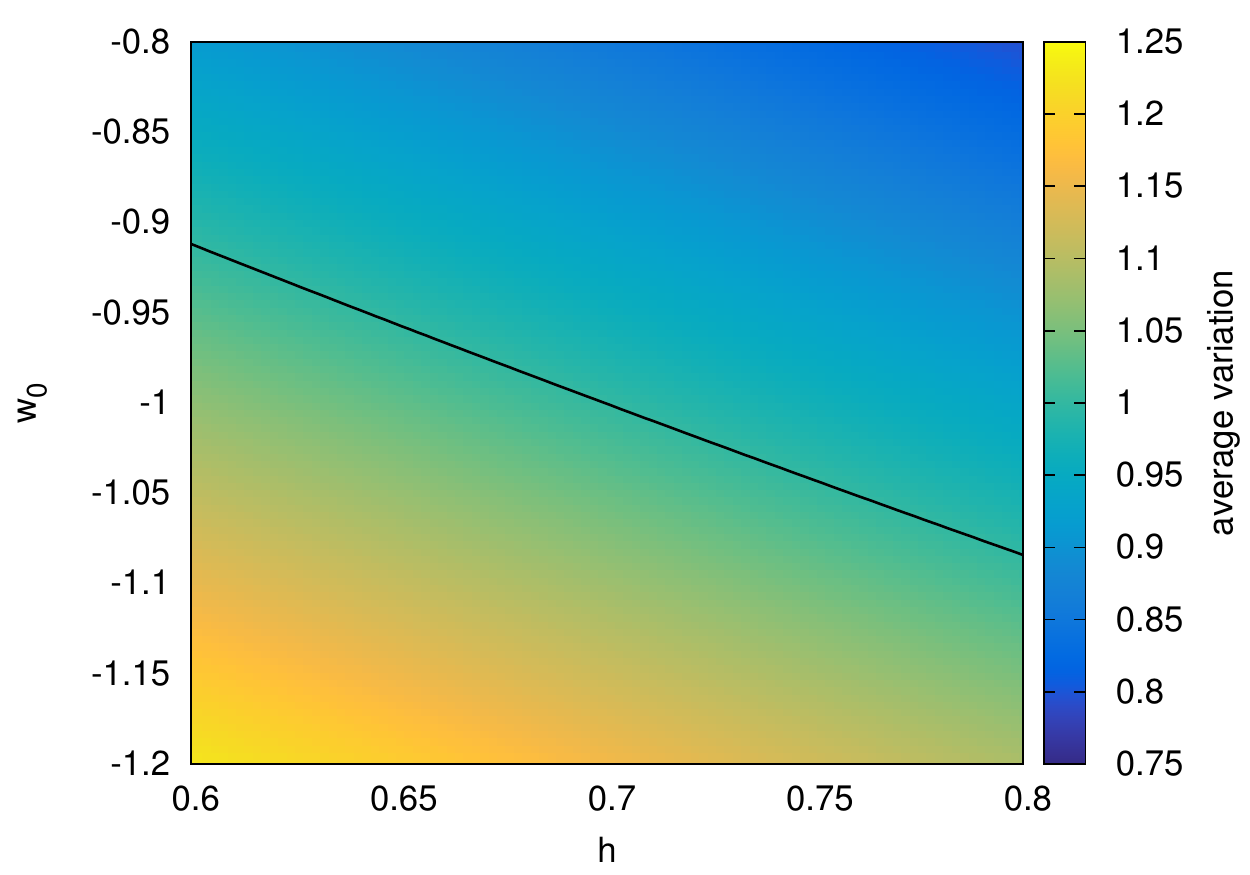}
\caption{Average variation of each component of the covariance matrix calculated by the relative value of the Frobenius norm of $C_{ij}$ with respect to the fiducial model. The black lines show the degeneracy of the frobenius norm along parameter space. In the ($\sigma_8,\Omega_\text{m})$ plane the degeneracy corresponds roughly to $\sigma_8\times\Omega_\text{m} = \text{const}$. In the $(w_\text{a},w_0)$ plane the black line shows models with constant effective equation of state. For $(h,n_\text{s})$ and $(w_0,h)$ we used fits of the form $hn_\text{s}^{a} =\text{const.}$ and $wh^b=\text{const.}$ respectively. We find $a \approx 2.75$ and $b \approx -0.6$. The covariance matrix is the same as in \autoref{Fig:2}. }
\label{Fig:3}
\end{center}
\end{figure*}

\subsection{Convergence spectrum covariance}
We apply our technique to the convergence spectrum, Eq. (\ref{eq:12}). As the convergence is a line-of-sight integral it will carry more information about the evolution of the Universe namely via the growth of structures and the geometrical evolution which both enter into Eq. (\ref{eq:12}); In particular we expect a much stronger variation of the weak lensing covariance matrix with the matter density $\Omega_m$. Technically, we use $N=60$ equidistant $\ell$-bins with width $\Delta\ell = 40$ and in the range from $\ell_1 =100$ to $\ell_N = 2500$. Furthermore, we assume a source redshift distribution with a mean redshift of 0.9, which would correspond to Euclid's anticipated redshift distribution. Since the redshift bins are summed over for each $\ell$ in the likelihood we only use one redshift bin. 

\autoref{Fig:2} shows the variation of the covariance matrix in parameter space, spanned by $\Omega_\text{m}$, $\sigma_8$, $h$, $n_\text{s}$ and the two dark energy parameters $w_0$ and $w_a$. Specifically, we quantify the isotropic and anisotropic changes of the covariance matrix by means of the trace and the determinant of a submatrix taken at low and high multipoles as before. 

In the $(\sigma_8,\Omega_\text{m})$-plane the isotropic magnification shows the usual degeneracy between these two parameters, because lensing is sensitive to the product of the two, to lowest order. As described before the anisotropic magnification shows an even stronger dependence on $\Omega_\text{m}$ due to different mode coupling contributions in the off-diagonal elements. 

The $(h,w_0)$-plane shows that the diagonal part of the covariance matrix is hardly influenced by the Hubble constant as its influence on the matter spectrum and the growth factor is rather small. In contrast, the equation of state parameter $w_0$ strongly influences the growth of structures and the geometry. In particular a more negative value of $w_0$ increases structure growth at early times and increases the lensing efficiency, thus leading to larger values for the lensing covariance. For the anisotropic magnification the dependence changes slightly due to the modification of the spectrum which becomes important in the non-Gaussian part of the covariance matrix. 

For the $(w_0,w_\text{a})$-plane, we adopted a linear evolution for the equation of state $w(a) = w_0 + w_\text{a}(1-a)$ \citep{Chevallier2001, Linder2006}. There are significant degeneracies between the two parameters for dark energy. This is due to the fact that the lensing signal depends on the equation of state function $w(a)$ through a triple integral, so effectively only on the average equation of state parameter. This effect can be seen in the anisotropic and isotropic magnification, thus mode coupling as well as the Gaussian part of the covariance matrix contain the same degeneracy.

\begin{figure*}
\begin{center}
\includegraphics[width = 0.45\textwidth]{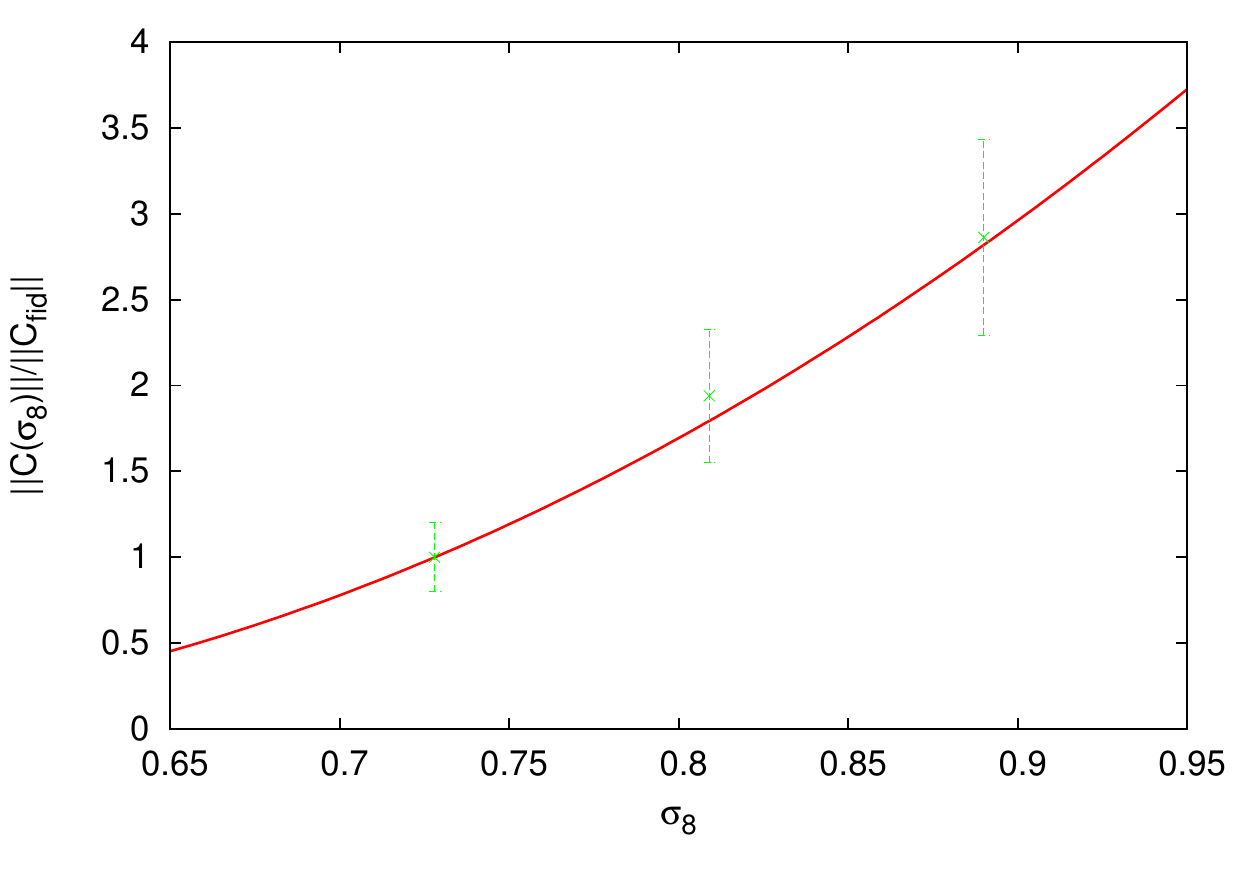}
\includegraphics[width = 0.45\textwidth]{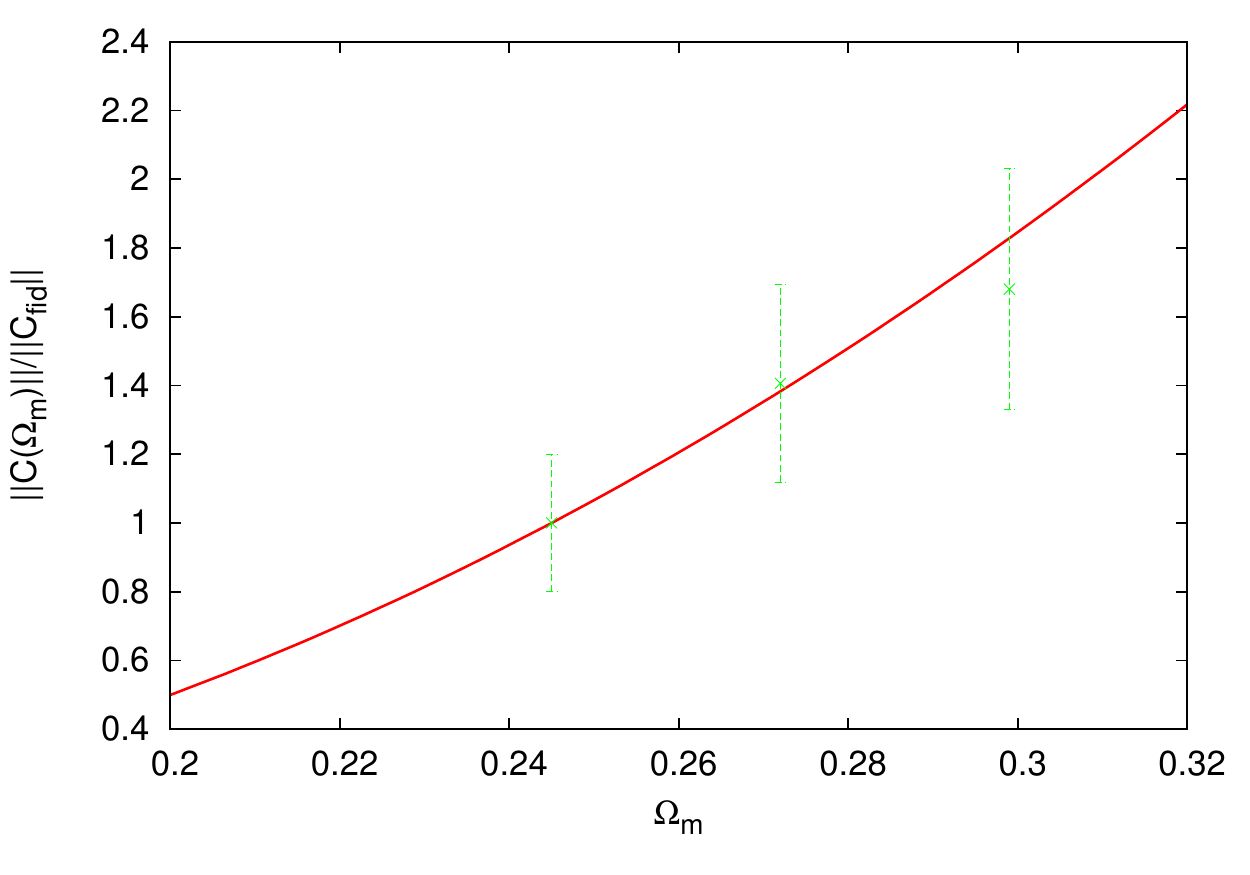}
\caption{Theoretical prediction (red line) vs. simulation (green crosses with error bars) of the weak lensing covariance matrix as a function of $\sigma_8$ (left) and with $\Omega_\text{m}$ (right). We compute the Frobenius-norm of the full covariance matrix and normalise it to the norm at the fiducial values $\sigma_8 = 0.728$ and $\Omega_\text{m} = 0.245$. The error bars indicate the variance within the set of numerical simulations.}
\label{Fig:4}
\end{center}
\end{figure*}

In order to get an intuition for the average change the covariance matrix while moving along parameter space we compare the Frobenius norm given by
\begin{equation}\label{eq:Forbnorm}
||C|| \equiv \sqrt{C_{ij}C_{ji}},
\end{equation}
relative to that of the fiducial cosmology. The ratio $||C(x^\prime_\alpha)||/||C(x_\alpha)||$ corresponds to the relative change of the covariance matrix as a function of a cosmological parameter $x_\alpha$. Nonetheless, it is clear that certain components of $C$ will change more drastically when changing the cosmology than others. \autoref{Fig:3} shows the average variation of the covariance matrix $C$ along different parameter combinations. The black line indicates an expected or fitted degeneracy of $C$. As already mentioned before lensing is sensitive to the product of $\sigma_8$ and $\Omega_\text{m}$, consequently the line plotted in the top left panel of \autoref{Fig:3} has $\Omega_\text{m}\times\sigma_8=\text{const}$ in the $(\Omega_\text{m},\sigma_8)$ plane. Clearly the expectation is well represented in the colour plot. For the dark energy equation of state in the top right panel we plot a line where the effective equation of state is $w_\text{eff} = -1$, with 
\begin{equation}
w_\text{eff} = \int_0^1\dd a\; w(a).
\end{equation}
In contrast, there are no straightforward arguments for the degeneracies regarding $(h,n_\text{s})$ and $(h,w_0)$. Therefore, we fit the degeneracy with  a power law of the form $xy^a = \text{const}$ and interpret the results.

For ($h,n_\text{s})$ we obtain an exponent of roughly $a=2.75$ with $hn_\text{s}^a=\text{const}$ which shows that the dependence on the spectral index is stronger than the dependence on the Hubble constant. Increasing $h$ shifts the peak of the matter spectrum to higher $k$. Thus, the amplitude of the spectrum becomes smaller if $\sigma_8$ is kept fixed, and as a consequence the values of the covariance matrix become smaller. Finally, the $(w_0,h)$ plane shows a degeneracy with $w_0h^a=\text{const}$, with $a \approx-0.6$. If $w_0>-1$ structure growth is decreased at early times, thus leading to smaller entries in the covariance matrix.

The analysis shows that the variation in the covariance matrix is strongest for the cosmological parameters responsible for structure formation.  Relative changes with respect to the fiducial model of roughly $80\%$ can occur in these parameters in the range of parameter values considered. The amount of variation is reduced if one applies priors and restricts the allowed parameter space, for instance by using cosmic microwave background data \citep[e.g.][]{Planck2015_XIII}.

It is important to note that we only kept terms linear in the generator. Of course one can also consider more terms in the expansion of the transformation matrix $U_{ik}$ in Eq. (\ref{eq:20}), although one loses commutativity of the generators and has to keep track of this using the Baker-Campbell-Hausdorff formula \citep{Schafer2016}. However, if the changes of the covariance matrices are small enough, $\sim$10$\%$, this approximation is justified.

\subsection{Comparing the variation with simulations}\label{sec:sim}
We use the weak lensing light-cones described in Sect. \ref{sec:24} to construct the spectrum of the convergence field $\kappa$ in $N = 9$ logarithmically equidistant bins in the angular wave vector, between $\ell_\text{min} = 186$ and $\ell_\text{max} = 1345$. Lower multipoles will exhibit large fluctuations due to the size of the simulation volume, while higher multipoles contain a strong shot noise contribution and suffer from resolution limits. For more details we refer to \citet{Kiessling2011}. 

For each cosmological parameter set from table \ref{tab:sim} the covariance matrix is estimated as described in Eq. (\ref{eq:CovSim}) from the available set of statistically equivalent simulations. Due to the relatively small number of realizations (see table \ref{tab:sim}) the estimator for the covariance itself is rather noisy with a relative error of $\sim$14$\%$ due to Poisson noise and convergence may not have been reached yet \citep[for convergence of covariance matrix estimators we refer to][]{Sellentin2016,Petri2016}. Therefore it is not useful to compare single components of the covariance matrix and instead we again compare the average change of the covariance matrix by means of the Frobenius norm, Eq. (\ref{eq:Forbnorm}). Note that it would certainly be sensible to test the algorithm against a more robust estimate of the covariance matrix. This, however, would require significant computational resources since at least roughly $\sim$10$^4$ simulations are needed at each point in parameter space to get a more reliable estimate of the covariance matrix with errors at the percent level.

\autoref{Fig:4} shows the theoretical prediction compared with the simulation for the two parameters $\sigma_8$ and $\Omega_\text{m}$. Clearly the theoretical prediction matches the simulation quite well for both parameters. By fixing a fiducial value we force the covariance matrix to agree at one point in parameter space. This, however, is not the case (at least in the non-linear regime), as it has been shown by various authors \citep[e.g.][]{Cooray2001c} that the covariance matrix with a trispectrum correction from Eulerian perturbation theory underestimates the covariance in comparison to that found in numerical simulations. Nonetheless the results show that the scaling of the spectra, trispectra and covariances with cosmological parameters can be captured well perturbativly even at one-loop and tree-level, respectively, even though the absolute magnitude cannot be precisely calculated.

In order to analyze the variation of the covariance matrix more accurately, i.e. comparing single components of it, more realizations of the simulated convergence field are needed. Furthermore the theoretical part can be improved by adding more order in perturbation theory, using the halo model, or including additional terms such as the halo sample variance \citep{Cooray2001, Takada2004, Takada2007, Takada2009, Sato2009, Kayo2012} as well as super sample covariance \citep{Takada2013}.

However, in this paper we intended to show that the covariance matrix exhibits variations across parameter space, which can be well captured via the linear mapping introduced in Eq. (\ref{eq:linmap}). It is therefore sufficient to keep the transformation matrix $U$ up to linear order. Furthermore, this will also preserve the Abelian structure of the transformation group, as commutativity is destroyed when including non-linear terms in the transformation \citep{Schafer2016}. A similar question in this context is related to the validity of the linear approximation; Because the variation of the covariance matrix is captured well by the model we introduced in Sect. \ref{sec:2}, we can compare the Frobenius norm of Lie approximated covariance matrix to the exactly calculated covariance matrix (at tree-level). As soon as the deviation becomes larger than some error threshold, which is given by the necessary accuracy for the covariance matrix, a new Lie basis should be constructed at this point. Alternatively the sampling of the parameter space can also be constructed on the level of the generators $T_{ij\alpha}$. For each direction, $\alpha$, this describes an $N\times N$ matrix relating the covariance matrix at one parameter point to the covariance matrix at another parameter point. The matrix $U_{ij}$ given in Eq. (\ref{eq:20}) is a good approximation for the transformation as long as the first term dominates over the higher order ones. This implies that a new Lie basis should be constructed in direction $\alpha$ at point $x_\alpha^\prime$ as soon as $(x_\alpha^\prime-x_\alpha)T_{ij\alpha} \approx 1$.

\begin{figure}
\begin{center}
\includegraphics[width = 0.45\textwidth]{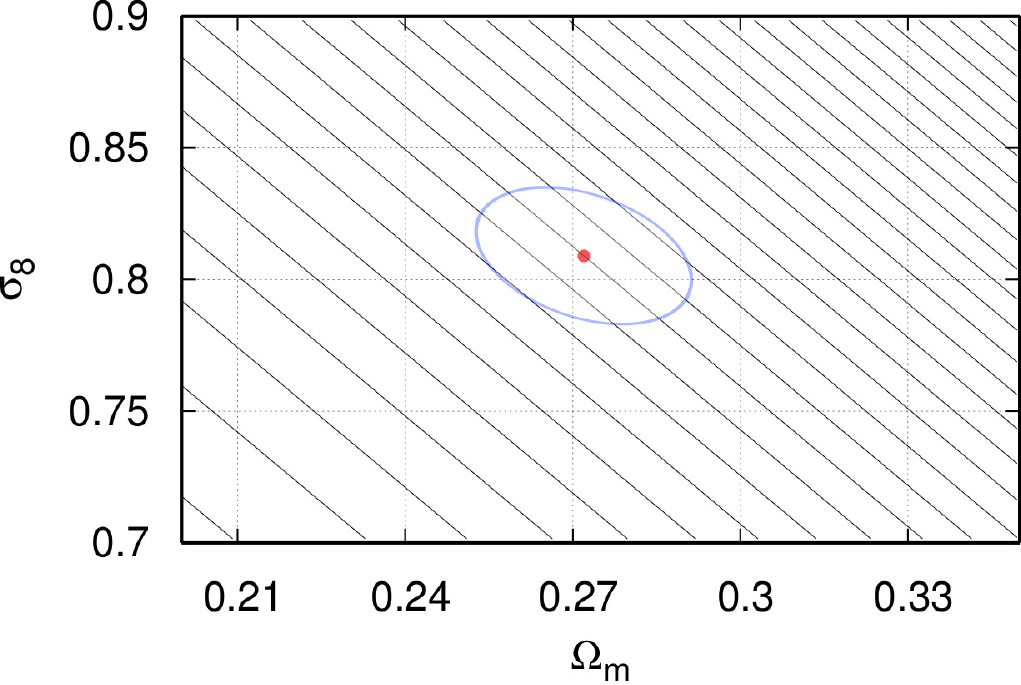}
\caption{Average change of the covariance matrix estimated using the Frobenius norm. The covariance matrix stays constant along the black lines and changes by $10\%$ from one black line to the other with respect to the fiducial model which is marked with a red dot. The blue ellipse indicates the approximate region of marginalized priors as found from covariance matrix measurements \citep{Planck2015_XIII}.}
\label{Fig:5}
\end{center}
\end{figure}

\autoref{Fig:5} shows one example for the outlined procedure. The Frobenius norm of the covariance matrix stays constant along the black lines, while it changes by roughly $10\%$ between neighbouring lines. The red dot marks the fiducial model and the blue ellipse indicates the marginalized priors from \citet{Planck2015_XIII} on both parameters. The direction of strongest change in the weak lensing covariance matrix is clearly into the direction of larger $\Omega_\text{m}$ and larger $\sigma_8$, while in the orthogonal direction the weak lensing covariance matrix does not change due to the proportionality of the weak lensing signal to the product $\Omega_\text{m}\times\sigma_8$. Consequently, it is sufficient to evaluate the covariance matrix by generating suites of simulations sparsely along lines of constant $\Omega_\text{m}\times\sigma_8$, while perpendicularly to that the variation of the weak lensing covariance must be followed in finer detail.

From another point of view, the eigenvectors of the matrix $A_{\alpha\beta}$, given by
\begin{equation}
A_{\alpha\beta} = \partial^2_{\alpha\beta} ||C||,
\end{equation}
with parameter directions $\alpha$ and $\beta$, point into the degeneracy direction and perpendicular to it, while the magnitude of the eigenvalues corresponds to the amount of change in these directions. For the completely degenerate case, as in \autoref{Fig:5}, the eigenvalue of the eigenvector, which is parallel to the degeneracy lines, would have eigenvalue zero. Accordingly, the grid on which the covariance is sampled could be rotated into the principal frame of matrix $A$, reducing this two dimensional sampling problem into a one dimensional one with degeneracy direction roughly given by the constraint $\Omega_\text{m}\times\sigma_8=\text{const}$. This procedure generalizes straightforwardly to higher dimensions.

\section{Conclusion}\label{sec:5}
In this paper we investigated variations of the covariance matrix of cosmic large-scale structure observations,  where non-linear structure formation processes generate non-Gaussian and non-diagonal contributions. We described the variation of the covariance matrix with a change of the cosmological model by constructing a basis, and considered as specific examples the matter density and weak lensing convergence power spectra. We worked with an analytical model for non-linear structure formation based on Eulerian perturbation theory and derived non-Gaussian contributions to the covariance matrices by evaluating the spectrum and the trispectrum in third order. This analytical model was juxtaposed with the results from numerical simulations, which showed that the fundamental scaling of the analytical model with the parameters $\Omega_\text{m}$ and $\sigma_8$ was reproduced correctly.

The covariance matrix of estimates of spectra depends on cosmological parameters, both in the linear and non-linear regime. We investigated the scaling of the covariance matrix with parameters from a $w$CDM-cosmology. By constructing a basis for the transformation which relates covariance matrices at different points in parameter space to each other we were able to predict the magnitude and degeneracies rather well, and were able to identify directions in parameter space associated with large changes in the covariance. 

Our formalism was able to represent variations in the covariance matrix for a wide region of the parameter space. In fact, it could describe variations much larger than that allowed by current experiments like Planck. Furthermore, the formalism also captured degeneracy lines, i.e. parameter combinations along which the covariance matrices effectively remain constant, which we showed to have clear physical explanations.

The identification of directions in parameter space in which the largest variations of the covariance matrix occur allows for an economical sampling with numerical simulations; This is feasible because our formalism effectively provides a metric which determines the distance in different directions in parameter space where the variation of the covariance matrix would be larger than a predefined threshold. Apart from predicting variations, our formalism is also well suited for inter- and extrapolation of covariance matrices which are ultimately determined from a large set of numerical simulations at discrete, specifically chosen, parameter points.

\section*{Acknowledgements}
RR acknowledges funding by the graduate college {\em Astrophysics of cosmological probes of gravity} by Landesgraduiertenakademie Baden-W{\"u}rttemberg. AK was supported in part by the Jet Propulsion Laboratory, run under contract by the California Institute of Technology for the National Aeronautics and Space Administration. AK was also supported in part by NASA ROSES 13-ATP13-0019.

\bibliographystyle{mnras}
\bibliography{MyBiB.bib,old_MasterBib.bib}

\bsp

\label{lastpage}

\end{document}